\documentclass[aps,prd,floats,nofootinbib,amssymb,
preprint]{revtex4-1}
\usepackage{graphicx}

\begin{document}

%-----------------------------------
% Title
%-----------------------------------

\title{Higgs couplings in a model with triplets}

\author{Heather~E.~Logan}
\email{logan@physics.carleton.ca}
\affiliation{Ottawa-Carleton Institute for Physics, Carleton University,
Ottawa K1S 5B6 Canada}

\author{Marc-Andr{\'e} Roy}
\affiliation{Ottawa-Carleton Institute for Physics, Carleton University,
Ottawa K1S 5B6 Canada}

\date{November 30, 2010}

\begin{abstract}
We study the couplings of a CP-even neutral Higgs boson $h$ in a model
containing one scalar SU(2)$_L$ doublet, one real triplet, and one
complex triplet with hypercharge 1.  Because the two triplets
contribute to the $\rho$ parameter with opposite signs, the triplet
vacuum expectation values can be sizable.  We show that (i) the $h WW$
and $h ZZ$ couplings can be larger than the corresponding values in
the Standard Model, and (ii) the ratio of the $WW$ and $ZZ$ couplings
of $h$ can be different than the corresponding ratio in the Standard
Model.  Neither of these results can occur in models containing only
Higgs doublets.  We also compute the rates for $gg \to h \to WW$ and
$gg \to h \to ZZ$ and find that, for reasonable parameter values and
$M_h \sim 140$--180~GeV, the hadron collider rate for $gg \to h \to
WW$ ($ZZ$) can be up to 20\% (5 times) larger than in the Standard
Model.  We discuss implications for Higgs coupling extraction at the
LHC.
\end{abstract}

\thispagestyle{empty}
\maketitle

%=======================================================================
% BEGIN MAIN TEXT
%=======================================================================
\section{Introduction}

The primary purpose of the CERN Large Hadron Collider (LHC) is to shed
light on the dynamics of electroweak symmetry breaking.  In the
Standard Model (SM), this involves the discovery of the Higgs boson
and the measurement of its mass-generating couplings to fermions and
gauge bosons.  Most extensions of the SM contain one or more
Higgs-like scalars that play a similar role in mass generation; the
characterization of such models also involves measurement of the
couplings of these scalars to the SM fermions and gauge bosons.

The LHC will be able to measure Higgs signal rates in a variety of
production and decay modes~\cite{ATLAS-CSC,CMS}, especially if
the Higgs mass lies in the range 114~GeV~$< M_h \lesssim 200$~GeV, as
suggested by direct searches~\cite{Barate:2003sz} and electroweak
precision data~\cite{EWPfit}.  These signal rate measurements will
provide access to the combinations of Higgs couplings that determine
the production cross sections and decay branching ratios involved in
each channel.  By taking ratios of Higgs signal rates with either the
production mode or the decay mode in common, ratios of
Higgs couplings-squared can be found with no model-dependent
assumptions~\cite{Zeppenfeld:2000td,Belyaev:2002ua}.

Unfolding the individual Higgs couplings from these rate measurements,
however, is a major challenge.  The main difficulty lies in the fact
that the LHC will not provide an absolute Higgs production cross
section measurement in any channel.  Together with the possible
presence of undetectable Higgs decay modes (e.g., $h \to gg$ or decays
into light quarks), this leads to a degeneracy in the Higgs coupling
fit corresponding to increasing all production couplings by a common
factor while simultaneously decreasing the detectable branching
fractions by the same factor.  This forces the adoption of theoretical
assumptions in Higgs coupling fits.  Analyses of LHC Higgs coupling
extraction prospects have dealt with this difficulty either by
assuming that the total Higgs width is dominated by the known SM decay
modes~\cite{Zeppenfeld:2000td,Belyaev:2002ua}, or by assuming that the
couplings of the Higgs to $WW$ and $ZZ$ are bounded from above by
their SM values~\cite{Duhrssen-ATLAS,Duhrssen:2004cv,Lafaye:2009vr}
(the latter analyses make the simultaneous assumption that the ratio
of the $hWW$ and $hZZ$ couplings is the same as in the SM).  The
latter approach allows the total width of the Higgs to be bounded from
above by a rate measurement in a channel with Higgs production and
decay via the $hWW$ or $hZZ$ coupling (e.g., weak boson fusion with $h
\to WW$).  These two assumptions about the $hWW$ and $hZZ$
couplings are valid in any Higgs sector that contains only SU(2)$_L$
doublets and/or singlets.  Larger SU(2)$_L$ multiplets are
usually considered unimportant because their contribution to the
$hWW$ and $hZZ$ couplings are generally tightly constrained by the
experimental limits on the $\rho$ parameter.

In this paper we explore the possibility that these assumptions about
the $hWW$ and $hZZ$ couplings are violated.  To do so, we need a model
with at least one Higgs doublet (required to generate fermion masses),
together with at least one larger SU(2)$_L$ representation (to give
rise to group-theoretic coefficients in the $hWW$ and $hZZ$ couplings
different from those of a doublet).  Because the couplings of interest
arise from quartic terms in the Lagrangian in which one scalar field
is replaced by its vacuum expectation value (vev), we need the vev(s)
of the larger SU(2)$_L$ multiplet(s) to be non-negligible.

To this end we study a model containing one Higgs doublet and two
triplets first introduced by Georgi and Machacek~\cite{Georgi:1985nv}.
One triplet is real with hypercharge zero while the other is complex
with hypercharge 1.\footnote{We use the convention that the
  hypercharge of the SM Higgs doublet is 1/2.}  Because these two
triplets contribute to the $\rho$ parameter with opposite sign, their
vevs need not be small so long as they are chosen such that their
contributions to $\rho$ cancel; this is achieved naturally by imposing
a global ``custodial'' SU(2)
symmetry~\cite{Georgi:1985nv,Chanowitz:1985ug,Gunion:1989ci,Gunion:1990dt,HHG}.
(The required global symmetry is however broken by hypercharge, so it
is no longer exact once radiative corrections are
included~\cite{Gunion:1990dt}.)  The model contains three CP-even
neutral scalars.  We study one of these states, which we call $h$ and
parameterize as a general mixture of the three weak eigenstates.  We
find that the $hWW$ and $hZZ$ couplings can be significantly larger
than their SM values, and that the ratio of these couplings can be
significantly different from the ratio in the SM.

This paper is organized as follows.  In the next section we summarize
the model and present general expressions for the couplings of
interest.  In Sec.~\ref{sec:numerics} we study the range of $hWW$ and
$hZZ$ couplings allowed in the model, subject to experimental
constraints that set a lower bound on the doublet vev.  We also
consider the impact on Higgs production and decay in two of the most
promising early Higgs discovery channels at the LHC, $gg \to h \to WW$
and $gg \to h \to ZZ$.  In particular, we find that the $gg \to h \to
ZZ$ rate can be significantly enhanced compared to that in the SM, due
to a suppression of the $h \to WW$ branching fraction in some parts of
parameter space without a simultaneous suppression of $h \to ZZ$.  In
Sec.~\ref{sec:howtotell} we discuss ways to experimentally determine
whether the $hZZ$ and $hWW$ couplings have the same ratio as in the SM
and whether these couplings are larger than their SM values.  In
Sec.~\ref{sec:conclusions} we summarize our conclusions.  Some
computational details are given in the appendices.

%=======================================================================
\section{The model}
\label{sec:model}

The Georgi-Machacek model~\cite{Georgi:1985nv} contains a complex SU(2)$_L$
scalar doublet $\Phi$ with hypercharge $Y = Q - T^3 = 1/2$, a real
triplet $\Xi$ with $Y = 0$, and a complex triplet $X$ with $Y = 1$,
with components and vevs denoted by
\begin{equation}
  \Phi = \left( \begin{array}{c} \phi^+ \\ 
    \frac{1}{\sqrt{2}}
    (v_{\phi} + \phi^{0,r} + i \phi^{0,i}) \end{array} \right), \qquad
  \Xi = \left( \begin{array}{c} \xi^+ \\ v_{\xi} + \xi^0 \\ 
    \xi^- \end{array} \right), \qquad
  X = \left( \begin{array}{c} \chi^{++} \\ \chi^+ \\
    \frac{1}{\sqrt{2}}
    (v_{\chi} + \chi^{0,r} + i \chi^{0,i}) \end{array} \right).
\end{equation}
Starting from the canonically normalized covariant derivative terms in
the scalar Lagrangian, we obtain the $W$ and $Z$ boson masses and
their couplings to the CP-even neutral scalars as follows:
\begin{eqnarray}
  \mathcal{L} &\supset& |\mathcal{D}_{\mu} \Phi |^2 
  + \frac{1}{2} |\mathcal{D}_{\mu} \Xi |^2
  + | \mathcal{D}_{\mu} X |^2 \nonumber \\
  & \supset & (v_{\phi} + \phi^{0,r})^2 
  \left[ \frac{g^2}{4} W_{\mu}^+ W^{- \mu} 
    + \frac{g^2 + g^{\prime 2}}{8} Z_{\mu} Z^{\mu} \right] \nonumber \\
  && + (v_{\xi} + \xi^0)^2 
  \left[4 \, \frac{g^2}{4} W_{\mu}^+ W^{- \mu} \right] \nonumber \\
  && + (v_{\chi} + \chi^{0,r})^2
  \left[2 \, \frac{g^2}{4} W_{\mu}^+ W^{- \mu}
    + 4 \, \frac{g^2 + g^{\prime 2}}{8} Z_{\mu} Z^{\mu} \right],
  \label{eq:kineticterms}
\end{eqnarray}
where $g$ and $g^{\prime}$ are the SU(2)$_L$ and U(1)$_Y$ gauge
couplings of the SM, respectively.  The covariant derivative and the
SU(2) generators for the triplet representation are given in
Appendix~\ref{app:covar}.  The squared masses of the $W$ and $Z$
bosons are then given by
\begin{equation}
  M_W^2 = \frac{g^2}{4}(v_{\phi}^2 + 4 v_{\xi}^2 + 2 v_{\chi}^2) 
  \equiv \frac{g^2}{4} v_{\rm SM}^2, 
  \qquad \qquad
  M_Z^2 = \frac{g^2 + g^{\prime 2}}{4}(v_{\phi}^2 + 4 v_{\chi}^2),
\end{equation}
where we define $v_{\rm SM}^2 = v_{\phi}^2 + 4 v_{\xi}^2 + 2
v_{\chi}^2 \simeq (246~{\rm GeV})^2$.

The different form of the triplet contributions to the $W$ and $Z$
masses results in a tree-level $\rho$ parameter different from 1,
\begin{equation}
  \rho \equiv \frac{M_W^2}{M_Z^2 \cos^2 \theta_W}
  = \frac{v_{\phi}^2 + 4 v_{\xi}^2 + 2 v_{\chi}^2}{v_{\phi}^2 + 4 v_{\chi}^2}
  = 1 + \frac{4 v_{\xi}^2 - 2 v_{\chi}^2}{v_{\phi}^2 + 4 v_{\chi}^2}
  \equiv 1 + \Delta \rho,
\end{equation}
where we used the definition $\cos \theta_W = g/\sqrt{g^2 + g^{\prime
    2}}$.  The $\rho$ parameter also receives radiative corrections in
the SM from isospin violation, depending mostly on the top quark and
SM Higgs masses.  After separating out the SM contributions, the
constraint on new sources of SU(2) breaking is given at the 2$\sigma$
level by~\cite{Amsler:2008zzb}
\begin{equation}
  \Delta \rho = 4^{+27}_{-7} \times 10^{-4} \qquad \qquad
  (2\sigma \ {\rm constraint}).
\end{equation}
This leads to the requirement $v_{\xi} \simeq v_{\chi}/\sqrt{2}$ to
high precision.  This relation can be enforced by imposing an
SU(2)$_R$ global symmetry on the Higgs potential, resulting in a
residual global ``custodial'' SU(2) symmetry---the diagonal subgroup
of SU(2)$_L \times$~SU(2)$_R$---after electroweak symmetry
breaking~\cite{Georgi:1985nv,Chanowitz:1985ug,Gunion:1989ci}.  The
SU(2)$_R$ global symmetry is broken by hypercharge (for the scalars,
hypercharge corresponds to the $T^3$ generator of SU(2)$_R$), so that
SU(2)$_R$-violating counterterms must be introduced at one-loop
level~\cite{Gunion:1990dt}.  Nevertheless, if the scale where
SU(2)$_R$ is a good symmetry is not too high, the approximate global
symmetry protects the $\rho$ parameter and has been used, e.g., to
control the effects of triplets on electroweak precision observables
in little Higgs models~\cite{Chang:2003zn}.

The imposition of custodial SU(2) on the Higgs potential results in
Higgs mass eigenstates after electroweak symmetry breaking that lie in
multiplets of the custodial SU(2) with common
masses~\cite{Georgi:1985nv}.  In particular, the physical states form
two singlets $H_1^0$ and $H_1^{0 \prime}$, a 3-plet $H_3^{+,0,-}$, and
a 5-plet $H_5^{++,+,0,-,--}$ under custodial SU(2), with the CP-even
neutral states given by~\cite{Georgi:1985nv}
\begin{equation}
  H_1^0 = \phi^{0,r}, \qquad 
  H_1^{0 \prime} = \frac{1}{\sqrt{3}}(\xi^0 + \sqrt{2} \chi^{0,r}), \qquad
  H_5^0 = \frac{1}{\sqrt{3}}(\sqrt{2} \xi^0 - \chi^{0,r}).
  \label{eq:H1H5}
\end{equation}
If the custodial SU(2) is exact, the two singlets $H_1^0$ and $H_1^{0
  \prime}$ can mix, but $H_5^0$ must be a mass eigenstate.

In this framework, the couplings of the CP-even neutral states to $W$,
$Z$, and fermion pairs are given by~\cite{Georgi:1985nv,Gunion:1989ci}
\begin{eqnarray}
  \bar g_{H_1^0 WW} = c_H, &\qquad& \bar g_{H_1^0 ZZ} = c_H, 
  \qquad \bar g_{H_1^0 ff} = \frac{1}{c_H}, \nonumber \\
  \bar g_{H_1^{0 \prime} WW} = \frac{2 \sqrt{2}}{\sqrt{3}} s_H, &\qquad&
  \bar g_{H_1^{0 \prime} ZZ} = \frac{2 \sqrt{2}}{\sqrt{3}} s_H, 
  \qquad \bar g_{H_1^{0 \prime} ff} = 0, \nonumber \\
  \bar g_{H_5^0 WW} = \frac{1}{\sqrt{3}} s_H, 
  &\qquad& \bar g_{H_5^0 ZZ} = -\frac{2}{\sqrt{3}} s_H,
  \qquad \bar g_{H_5^0 ff} = 0,
\end{eqnarray}
where we define $\cos \theta_H \equiv c_H = v_{\phi}/v_{\rm SM}$, $s_H
= \sqrt{1 - c_H^2}$, and the barred couplings are normalized to the
corresponding SM Higgs couplings, $\bar g_{hxx} \equiv
g_{hxx}/g_{H_{\rm SM} xx}$.

In this paper we relax the assumption of custodial SU(2) in the Higgs
potential.  We still require that $v_{\xi} \simeq v_{\chi}/\sqrt{2}$
in order to obey the constraint on the $\rho$ parameter, but we allow
arbitrary mixing of the three CP-even neutral Higgs states.  This is a
fine-tuned situation, but it allows us to illustrate the full range of
couplings allowed by experimental constraints.  We will also present
results for custodial SU(2)-preserving mixing.  The most general
gauge-invariant Higgs potential (without custodial SU(2)) is given in
Appendix~\ref{app:potential}, where we show that enough parameter
freedom exists to allow arbitrary mixing even after the vevs are fixed
as above.

The doublet and triplet vevs can be parameterized in terms of $v_{\rm
  SM}$, $c_H$, and $\Delta \rho$ according to
\begin{eqnarray}
  v_{\phi}^2 &=& c_H^2 v_{\rm SM}^2, \nonumber \\
  v_{\xi}^2 &=& \frac{1}{8} 
  \left[s_H^2 + \frac{\Delta \rho}{1 + \Delta \rho} \right] 
  v_{\rm SM}^2 \simeq \frac{1}{8} s_H^2 v_{\rm SM}^2, \nonumber \\
  v_{\chi}^2 &=& \frac{1}{4} 
  \left[s_H^2 - \frac{\Delta \rho}{1 + \Delta \rho} \right] v_{\rm SM}^2
  \simeq \frac{1}{4} s_H^2 v_{\rm SM}^2.
  \label{eq:vevs}
\end{eqnarray}

We consider two parameterizations of the mixing among the three neutral
states.  First, we parameterize the CP-even neutral Higgs state $h$ of 
interest as
\begin{equation}
  h = a \, \phi^{0,r} + b \, \xi^0 + c \, \chi^{0,r},
  \label{eq:abc}
\end{equation}
where $a^2 + b^2 + c^2 = 1$.\footnote{In this notation, $H_1^0$,
  $H_1^{0 \prime}$, and $H_5^0$ correspond to $(a,b,c) = (1,0,0)$,
  $(0,\sqrt{1/3},\sqrt{2/3})$, and $(0,\sqrt{2/3},-\sqrt{1/3})$,
  respectively.}  The couplings of $h$ are then given by
\begin{equation}
  \bar g_{hWW} = \frac{a v_{\phi} + 4b v_{\xi} + 2c v_{\chi}}{v_{\rm SM}},
  \qquad \bar g_{hZZ} = \frac{a v_{\phi} + 4c v_{\chi}}{v_{\rm SM}},
  \qquad \bar g_{hff} = \frac{a v_{\rm SM}}{v_{\phi}} = \frac{a}{c_H}.
\end{equation}
In the limit $\Delta \rho \to 0$, the $hWW$ and $hZZ$ couplings can be
written as
\begin{equation}
  \bar g_{hWW} = a c_H + (\sqrt{2} b + c) s_H,
  \qquad \bar g_{hZZ} = a c_H + 2 c s_H.
  \label{eq:coupsdrzero}
\end{equation}

As a second parameterization we can write
\begin{equation}
  h = \cos\phi \, ( \cos\theta \, H_1^0 + \sin\theta \, H_1^{0 \prime})
  + \sin\phi \, H_5^0,
\end{equation}
where $\theta$ parameterizes the mixing between the two custodial
SU(2) singlets and $\phi$ parameterizes the mixing between the
custodial SU(2) singlets and the 5-plet.  These mixing
angles are related to $a$, $b$, and $c$ above by
\begin{equation}
  a = \cos\phi \cos\theta, \qquad
  b = \frac{1}{\sqrt{3}} \cos\phi \sin\theta + \sqrt{\frac{2}{3}} \sin\phi, 
    \qquad
  c = \sqrt{\frac{2}{3}} \cos\phi \sin\theta - \frac{1}{\sqrt{3}} \sin\phi.
\end{equation}
In the limit $\Delta \rho = 0$ we have for the couplings,
\begin{eqnarray}
  \bar g_{hWW} &=& \cos\phi \cos\theta \, c_H 
  + \cos\phi \sin\theta \frac{2 \sqrt{2}}{\sqrt{3}} s_H
  + \sin\phi \frac{1}{\sqrt{3}} s_H, \nonumber \\
  \bar g_{hZZ} &=& \cos\phi \cos\theta \, c_H 
  + \cos\phi \sin\theta \frac{2 \sqrt{2}}{\sqrt{3}} s_H
  - \sin\phi \frac{2}{\sqrt{3}} s_H, \nonumber \\
  \bar g_{hff} &=& \cos\phi \cos\theta \frac{1}{c_H}.
  \label{eq:coupsthetaphi}
\end{eqnarray}

Though less compact, this second parameterization will be more useful
for our analysis.  We note in particular that when $\sin\phi = 0$, $h$
is a mixture of custodial SU(2) singlets only, and $\bar g_{hWW} =
\bar g_{hZZ} \equiv \bar g_{hVV}$.  For $\sin\phi = 0$ we have, 
\begin{equation}
  \bar g_{hVV} = \cos\theta \, c_H 
  + \sin\theta \frac{2 \sqrt{2}}{\sqrt{3}} s_H.
\end{equation}
This coupling reaches a maximum value of
\begin{equation}
  \bar g_{hVV}^{\rm max} = \sqrt{\frac{8 - 5 c_H^2}{3}}
  \qquad {\rm when} \qquad
  \sin\theta^{\rm max} = \sqrt{\frac{8 s_H^2}{8 - 5 c_H^2}}.
  \label{eq:thetamax}
\end{equation}

%=======================================================================
\section{Numerical results}
\label{sec:numerics}

Imposing $\Delta \rho = 0$, the largest $hWW$ and $hZZ$
couplings occur when $v_{\phi} = 0$ and $v_{\chi} = \sqrt{2} v_{\xi} =
v_{\rm SM}/2$.  This yields coupling maxima of $\bar g_{hWW} =
\sqrt{3}$ for $h = \sqrt{\frac{2}{3}} \xi^0 + \frac{1}{\sqrt{3}}
\chi^{0,r}$, for which $\bar g_{hZZ} = 2/\sqrt{3}$; and $\bar g_{hZZ}
= 2$ for $h = \chi^{0,r}$, for which $\bar g_{hWW} = 1$.  This
situation is clearly unrealistic, however, because a nonzero value of
$v_{\phi}$ is needed to generate fermion masses.  This leads to a
lower bound on $c_H$ (equivalently $v_{\phi}$) from processes that
constrain the top quark Yukawa coupling in multi-Higgs-doublet models.
In particular, the $\bar b b$ fraction in hadronic $Z$ decays, $R_b$,
was studied in the Georgi-Machacek model in Ref.~\cite{Haber:1999zh}.
For a custodial SU(2)-symmetric spectrum, $H_3^{\pm}$ contributes to
$R_b$ at one-loop level and leads to the constraint $c_H \geq 0.9$
(0.6) for $M_3 = 100$ (1000)~GeV.  If mixing between $H_3^{\pm}$ and
$H_5^{\pm}$ is allowed, new contributions arise that can have either
sign.  In what follows we plot results for $c_H = 0.9$ and 0.95 as
reasonable values.  We also show the $h$ couplings to $WW$ and $ZZ$
with $c_H = 0.6$ as an extreme case.

%=======================================================================
\subsection{Higgs couplings}

\begin{figure}
\resizebox{\textwidth}{!}{
\includegraphics{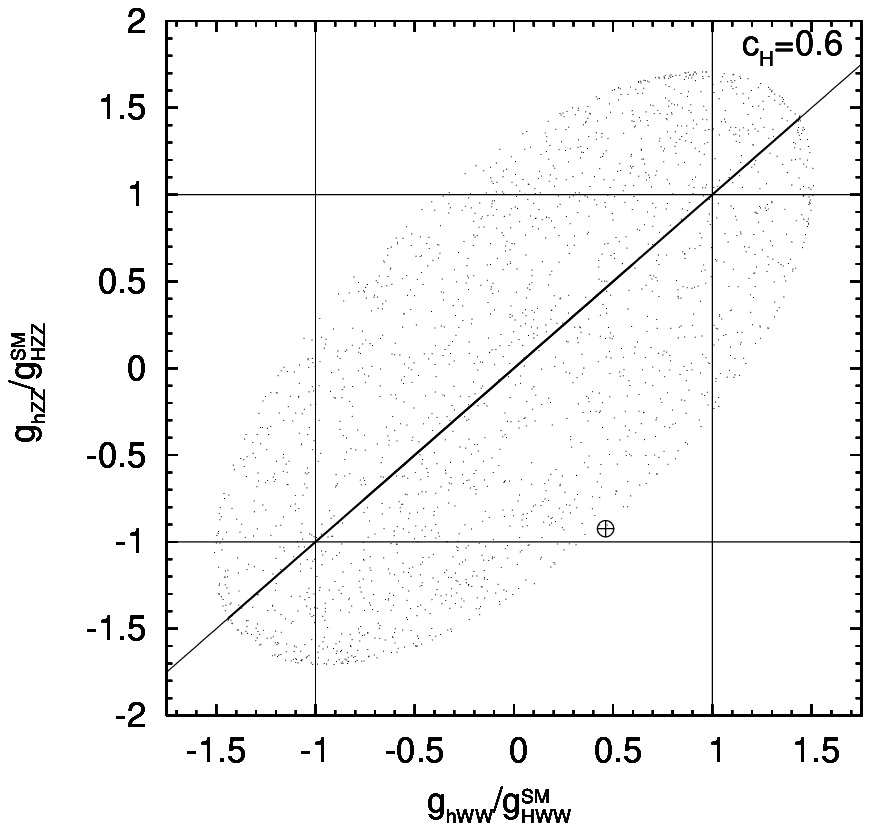}
\includegraphics{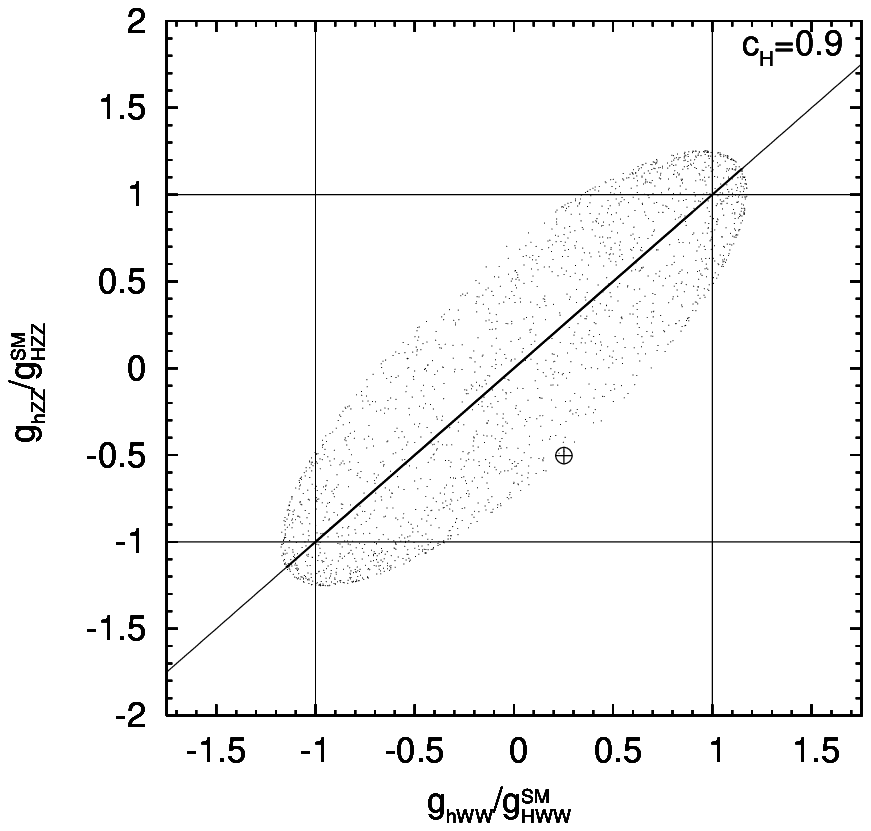}
\includegraphics{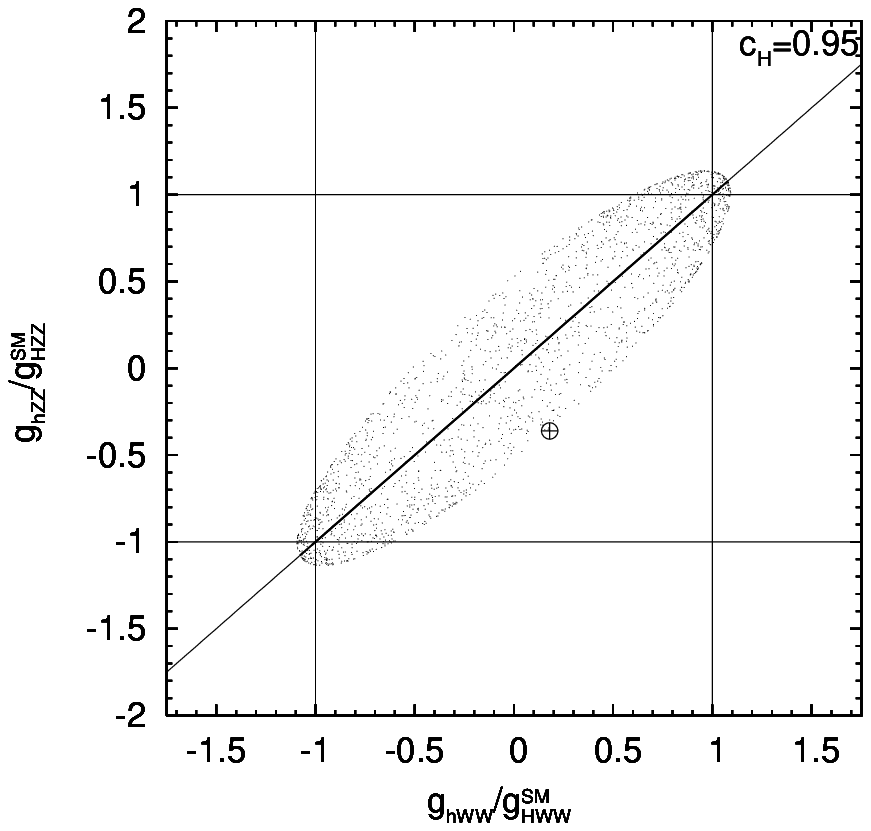}}
\caption{Normalized Higgs couplings $\bar g_{hZZ}$ versus $\bar
  g_{hWW}$ for a mixed state $h$ in the Georgi-Machacek model, for
  $c_H = 0.6$, 0.9, and 0.95 (left to right).  We sampled both signs
  for $a$, $b$, and $c$ in Eq.~(\ref{eq:coupsdrzero}).  Points
  corresponding to a mixture of $H_1^0$ and $H_1^{0\prime}$ lie along
  the diagonal $\bar g_{hWW} = \bar g_{hZZ}$.  $H_5^0$ is indicated by
  the crossed circle on the lower right boundary of the ellipse.}
\label{fig:ellipses}
\end{figure}

When plotted against each other for a fixed value of $c_H$, the
expressions for $\bar g_{hWW}$ and $\bar g_{hZZ}$ given in
Eqs.~(\ref{eq:coupsdrzero}) and (\ref{eq:coupsthetaphi}) describe a
filled ellipse centered at the origin as shown in
Fig.~\ref{fig:ellipses}.  The ellipse is bounded by a curve
corresponding to $\theta = \theta^{\rm max}$ as defined in
Eq.~(\ref{eq:thetamax}) and given parametrically by
\begin{eqnarray}
  \bar g_{hWW} &=& \cos\phi \sqrt{\frac{8 - 5 c_H^2}{3}}
  + \sin\phi \frac{s_H}{\sqrt{3}}, \nonumber \\
  \bar g_{hZZ} &=& \cos\phi \sqrt{\frac{8 - 5 c_H^2}{3}}
  - \sin\phi \frac{2 s_H}{\sqrt{3}}.
\end{eqnarray}
The bounding ellipse can be conveniently expressed by defining
$g_{\pm} \equiv (\bar g_{hZZ} \pm \bar g_{hWW})/\sqrt{2}$; we then
obtain
\begin{equation}
  1 = \frac{3}{2 (8 - 5 c_H^2)} g_+^2
  - \frac{1}{(8 - 5 c_H^2)} g_+ g_-
  + \frac{(11 - 7 c_H^2)}{2 s_H^2 (8 - 5 c_H^2)} g_-^2.
\end{equation}
Note that this ellipse is not oriented along the diagonal $\bar
g_{hWW} = \bar g_{hZZ}$, but instead has its semimajor axis tilted at
a small angle $\varphi$ above the diagonal, where
\begin{equation}
  \tan 2 \varphi = \frac{s_H^2}{2 (1 + s_H^2)}.
\end{equation}

In the custodial SU(2)-symmetric case, $H_5^0$ must be a mass
eigenstate but $H_1^0$ and $H_1^{0 \prime}$ can mix.  Points
corresponding to $h = \cos\theta H_1^0 + \sin\theta H_1^{0 \prime}$
fall along the diagonal, $\bar g_{hWW} = \bar g_{hZZ}$.  The
normalized couplings of $H_5^0$ are given in Table~\ref{tab:H5coups}.
In particular, $\bar g_{H_5^0 ZZ} = -2 \bar g_{H_5^0 WW}$.  The $h =
H_5^0$ point falls on the boundary of the ellipses in
Fig.~\ref{fig:ellipses}, marked with a crossed circle.  

\begin{table}
\begin{tabular}{ccc}
\hline \hline
$c_H$ & $\bar g_{H_5^0 WW}$ & $\bar g_{H_5^0 ZZ}$ \\
\hline
0.6 & 0.462 & $-0.924$ \\
0.9 & 0.252 & $-0.503$ \\
0.95 & 0.180 & $-0.361$ \\
\hline \hline
\end{tabular}
\caption{Normalized couplings of $H_5^0$ to $WW$ and $ZZ$ for various
  values of $c_H$.}
\label{tab:H5coups}
\end{table}

%=======================================================================
\subsection{Higgs production and decay rates}

We now compute the rates for $gg \to h \to WW$ and $gg \to h \to ZZ$,
normalized to their SM values.  In the SM, these processes constitute
the most promising discovery modes for an intermediate-mass Higgs.
Our calculation is valid for any hadron collider, because dependence
on the initial-state parton densities and collider center-of-mass
energy cancel in the ratio with the SM rate.  We assume throughout
that the narrow-width approximation can be applied.  For $VV = WW$ 
or $ZZ$ we have,
\begin{equation}
  \sigma(gg \to h \to VV)/\sigma_{\rm SM}
  \equiv \frac{\sigma(gg \to h \to VV)}{\sigma_{\rm SM}(gg \to H \to VV)}
  = \frac{\sigma(gg \to h)}{\sigma_{\rm SM}(gg \to H)}
  \frac{{\rm BR}(h \to VV)}{{\rm BR_{\rm SM}}(H \to VV)}.
\end{equation}
The $gg \to h$ production process proceeds through the coupling of $h$
to a fermion loop and is dominated in the SM by the top quark
contribution.  Because all fermion couplings are scaled by the same
factor $\bar g_{hff}$ in the Georgi-Machacek model, we
have,\footnote{At two-loop order, there are electroweak contributions
  to the $gg \to h$ amplitude in which $h$ couples to $WW$ or $ZZ$
  inside the loop, which should instead be scaled by $\bar g_{hWW}$ or
  $\bar g_{hZZ}$, respectively.  In the SM, these contributions result
  in a 5--6\% positive correction to the cross
  section~\cite{Anastasiou:2008tj}.  Ignoring their different coupling
  dependence thus results in an error of order $0.06 \, \bar g_{hff}
  (\bar g_{hff} - \bar g_{hWW, hZZ})$ on the right-hand side of
  Eq.~(\ref{eq:sigmaggh}).}
\begin{equation}
  \frac{\sigma(gg \to h)}{\sigma_{\rm SM}(gg \to H)} = \bar g_{hff}^2.
  \label{eq:sigmaggh}
\end{equation}
To compute the decay branching ratios in the Georgi-Machacek model, we
start with the partial widths of the SM Higgs of the same mass,
computed using the public FORTRAN code {\tt HDECAY} version
3.531~\cite{HDECAY}.  The corresponding partial widths of $h$ are computed
as follows:
\begin{eqnarray}
  \Gamma(h \to WW) &=& \bar g_{hWW}^2 \Gamma_{\rm SM}(H \to WW), \qquad
  \Gamma(h \to ZZ) = \bar g_{hZZ}^2 \Gamma_{\rm SM}(H \to ZZ) \nonumber \\
  \Gamma(h \to f \bar f) &=& \bar g_{hff}^2 \Gamma_{\rm SM}(H \to f \bar f),
  \qquad
  \Gamma(h \to gg) = \bar g_{hff}^2 \Gamma_{\rm SM}(H \to gg),
\end{eqnarray}
where $f \bar f$ refers to any fermion pair.  The remaining partial
widths, $\Gamma(h \to \gamma \gamma)$ and $\Gamma(h \to \gamma Z)$,
constitute a very small portion of the $h$ total width; for simplicity
we ignore the corrections to these modes and instead take them equal
to the corresponding SM Higgs widths.\footnote{The error introduced by
  this assumption in the $h$ branching ratio calculation is at the
  level of $10^{-3} \, (1 - \bar g_{hff, hWW}^2)$.}  The branching
ratios of $h$ are then given by taking the ratio of the relevant
partial width to the $h$ total width.

\begin{figure}
\resizebox{\textwidth}{!}{
\includegraphics{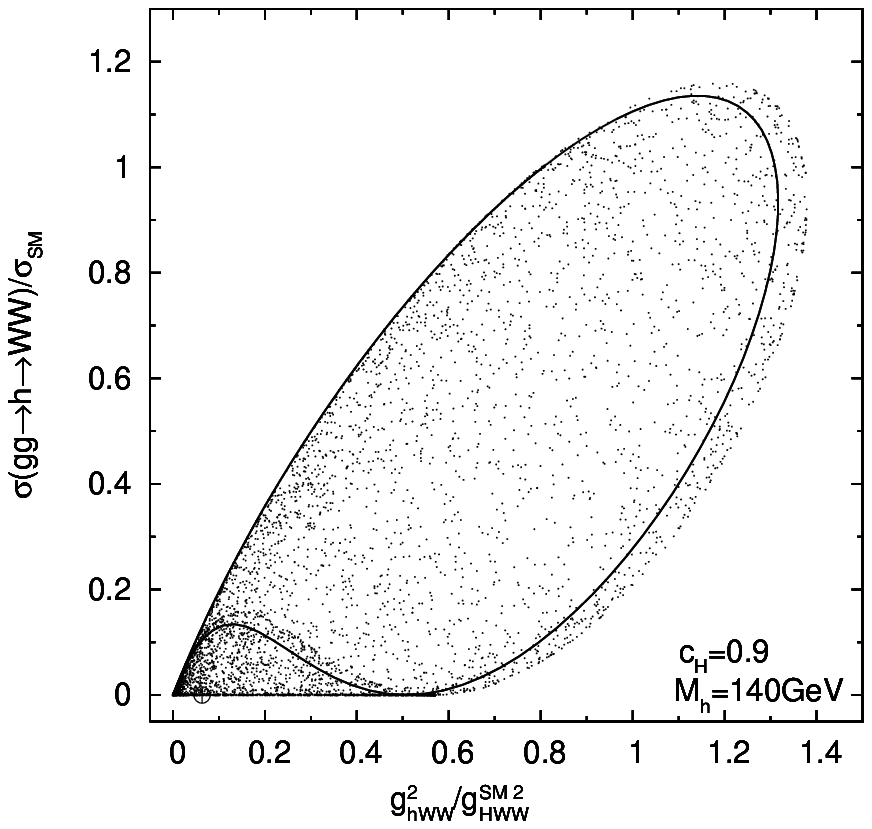}
\includegraphics{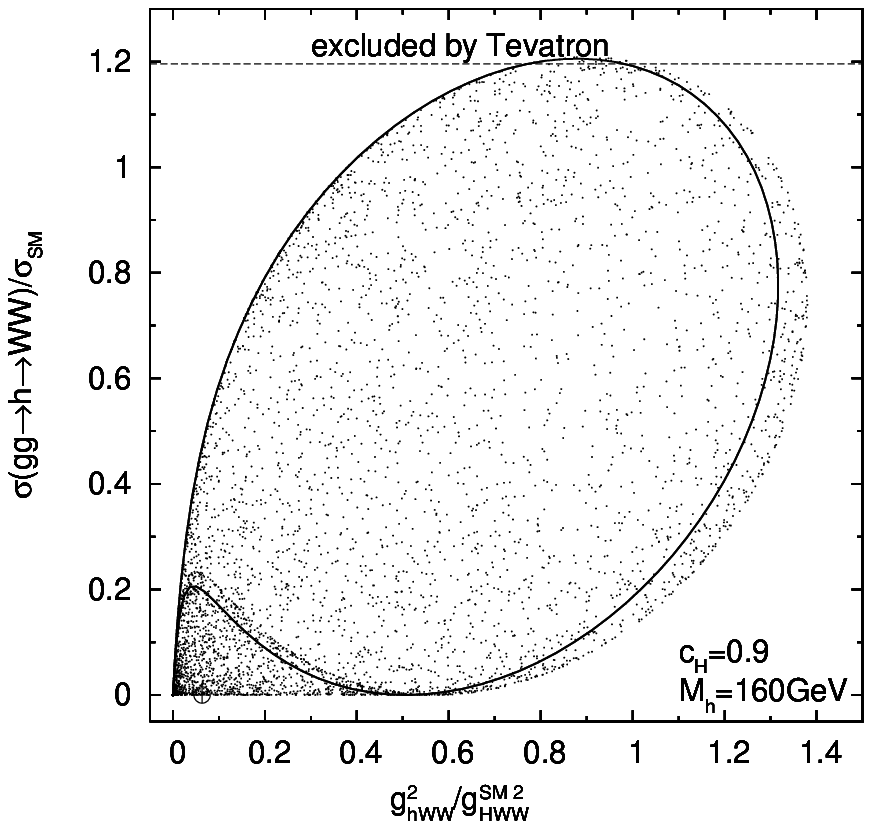}
\includegraphics{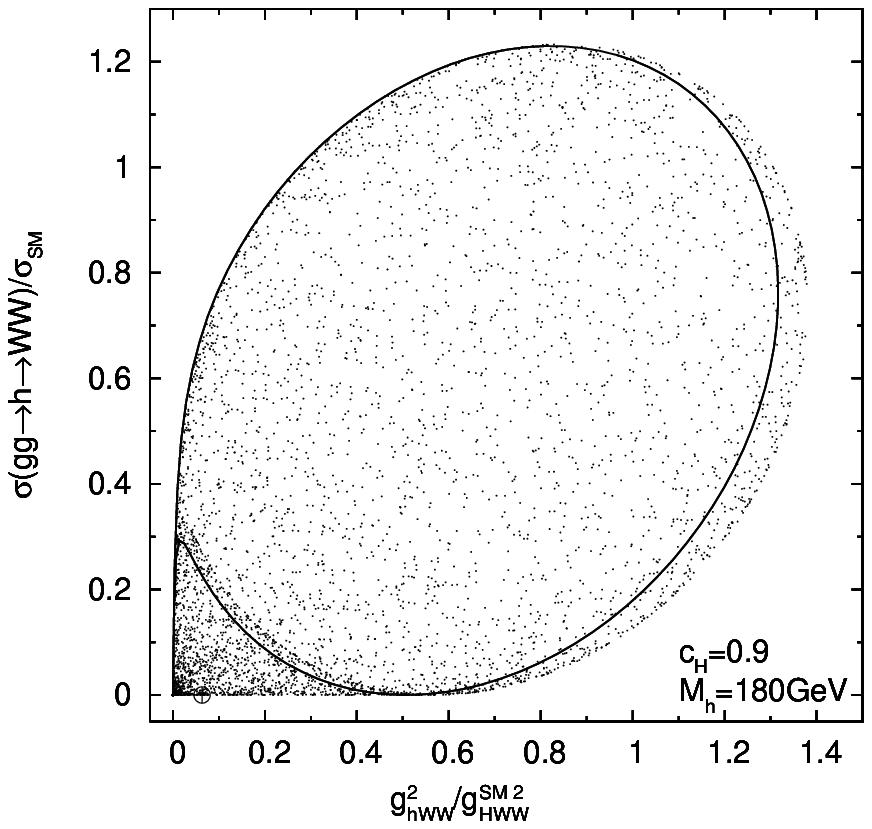}}
\caption{The rate for $gg \to h \to WW$ normalized to its SM value
  plotted as a function of $\bar g_{hWW}^2$, for $M_h = 140$, 160 and
  180~GeV (left to right), with $c_H = 0.9$.  The solid line shows
  points for which $h = \cos\theta H_1^0 + \sin\theta H_1^{0 \prime}$
  and the crossed circle near the origin indicates the point
  corresponding to $H_5^0$.  The dashed horizontal line shows the
  Tevatron upper limit on the rate for $gg \to h \to WW$ for the
  corresponding Higgs mass from Ref.~\cite{Aaltonen:2010sv}.}
\label{fig:W9W}
\end{figure}

\begin{figure}
\resizebox{\textwidth}{!}{
\includegraphics{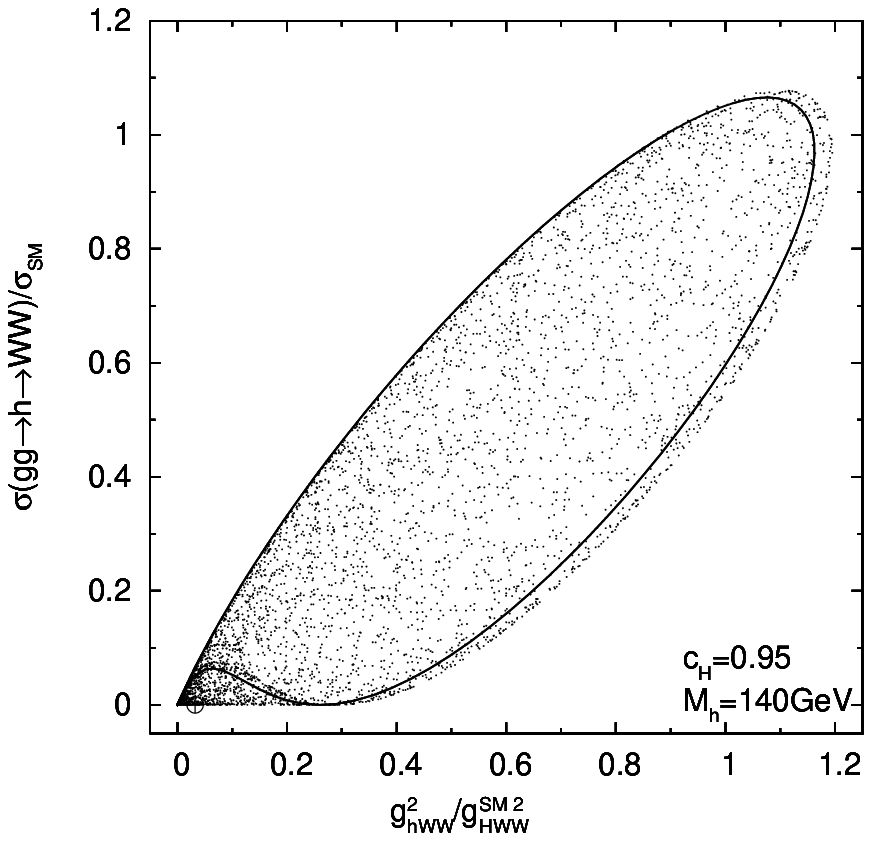}
\includegraphics{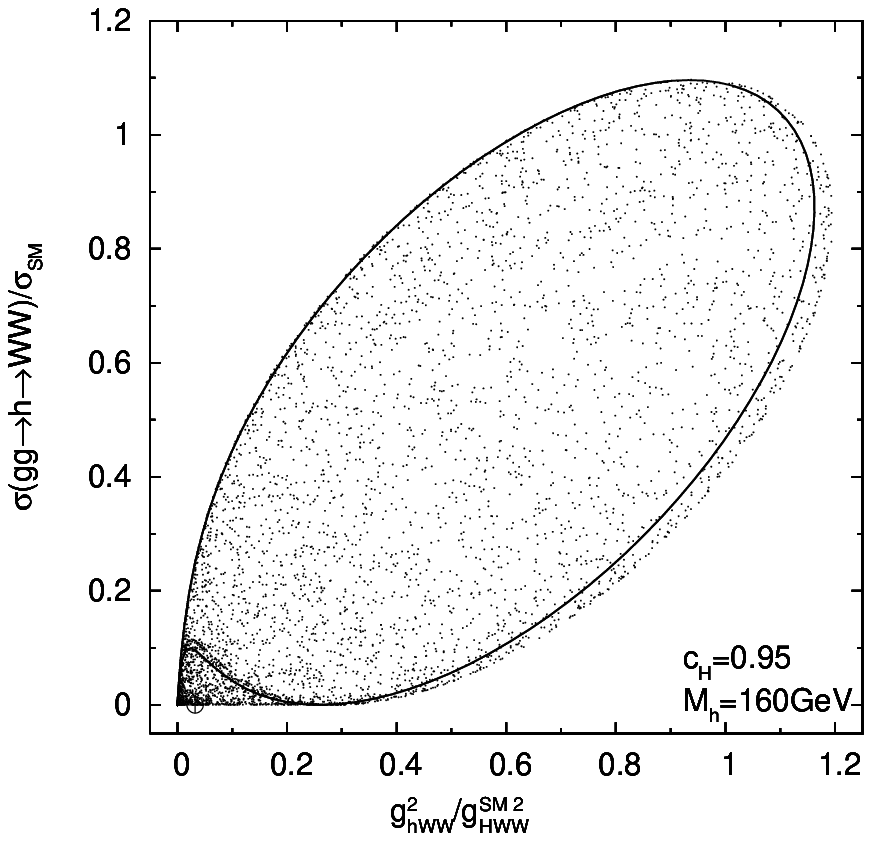}
\includegraphics{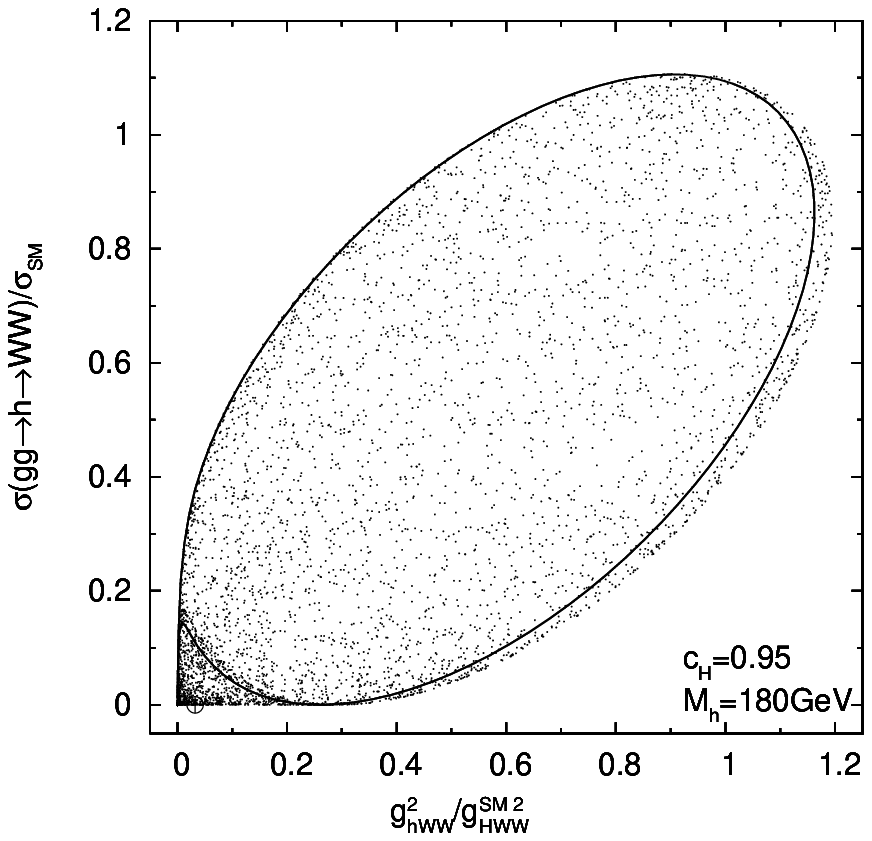}}
\caption{As in Fig.~\ref{fig:W9W} but for $c_H = 0.95$.}
\label{fig:W95W}
\end{figure}

\begin{figure}
\resizebox{\textwidth}{!}{
\includegraphics{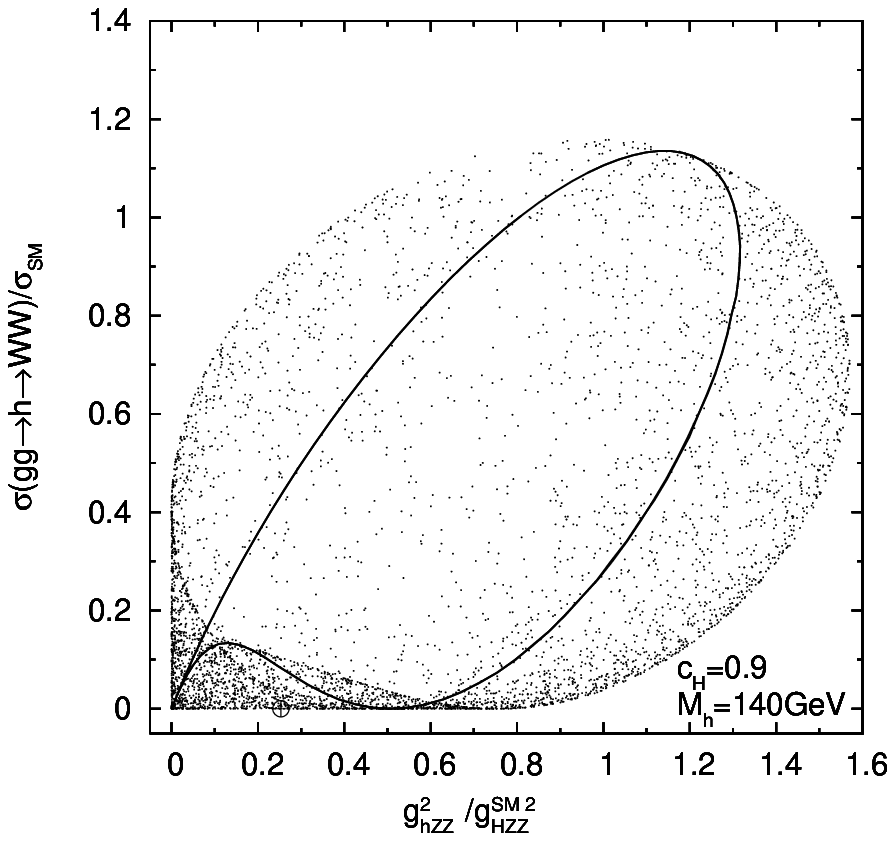}
\includegraphics{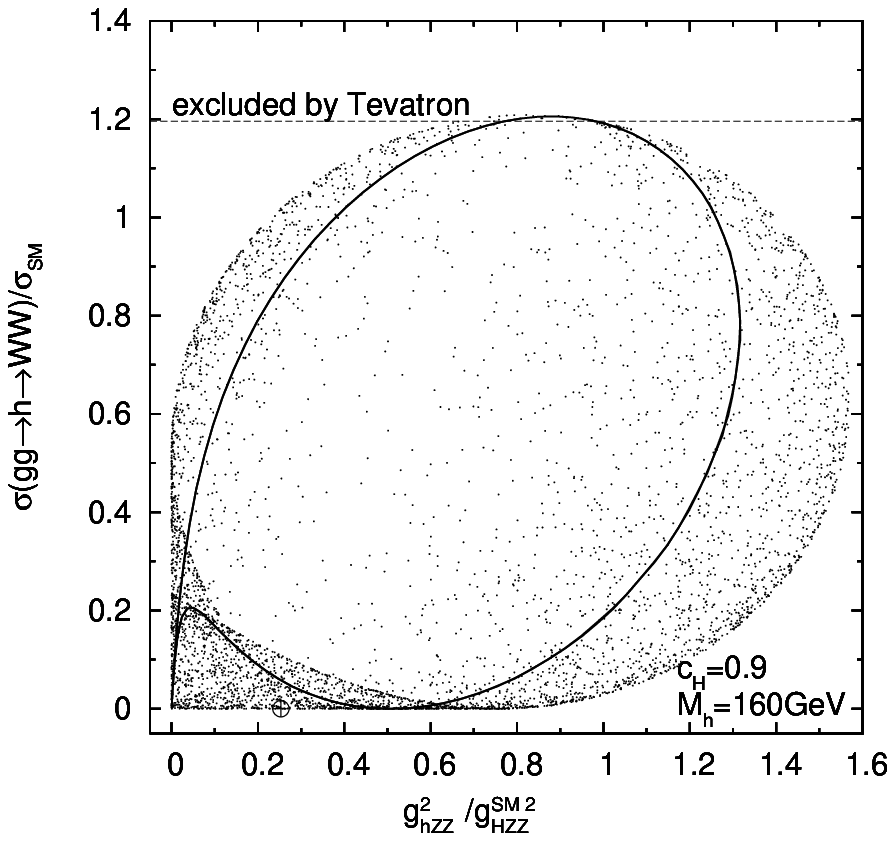}
\includegraphics{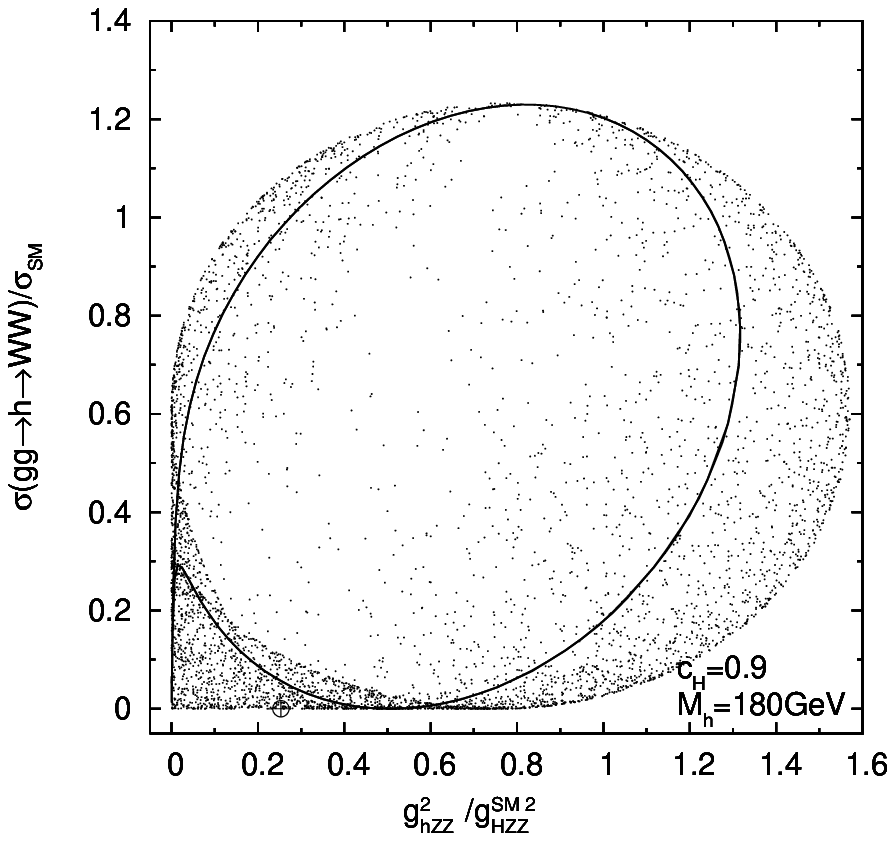}}
\caption{The rate for $gg \to h \to WW$ normalized to its SM value
  plotted as a function of $\bar g_{hZZ}^2$, for $M_h = 140$, 160 and
  180~GeV (left to right), with $c_H = 0.9$.  The solid line shows
  points for which $h = \cos\theta H_1^0 + \sin\theta H_1^{0 \prime}$
  and the crossed circle at the lower edge of the allowed region
  indicates the point corresponding to $H_5^0$.  The dashed horizontal
  line shows the Tevatron upper limit on the rate for $gg \to h \to
  WW$ for the corresponding Higgs mass from
  Ref.~\cite{Aaltonen:2010sv}.}
\label{fig:W9Z}
\end{figure}

\begin{figure}
\resizebox{\textwidth}{!}{
\includegraphics{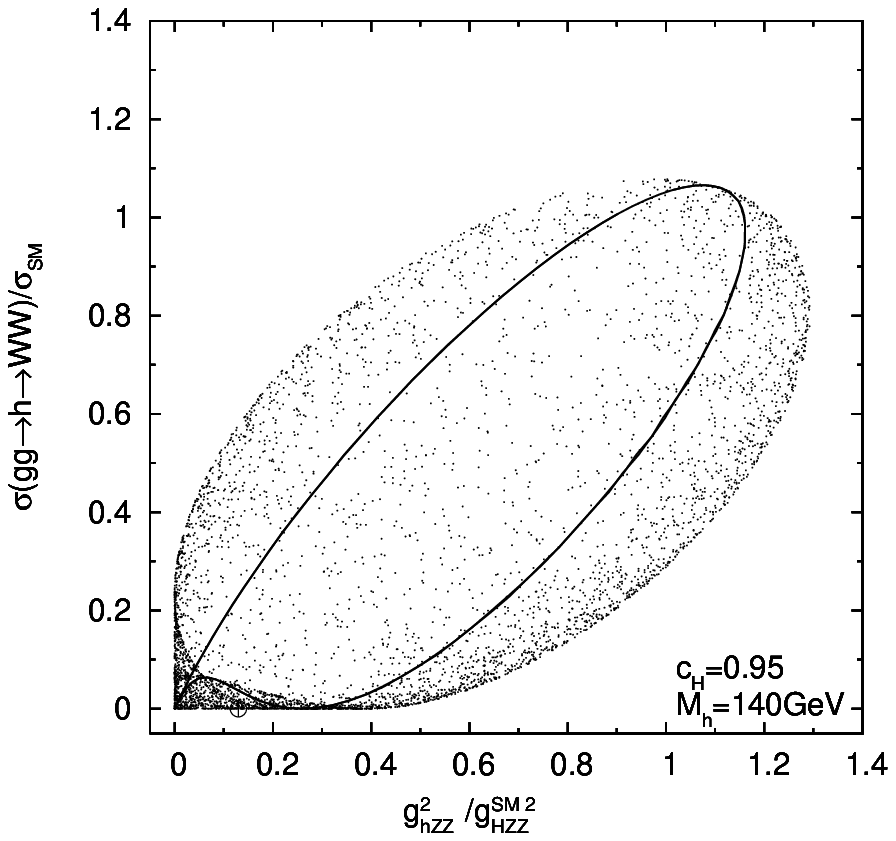}
\includegraphics{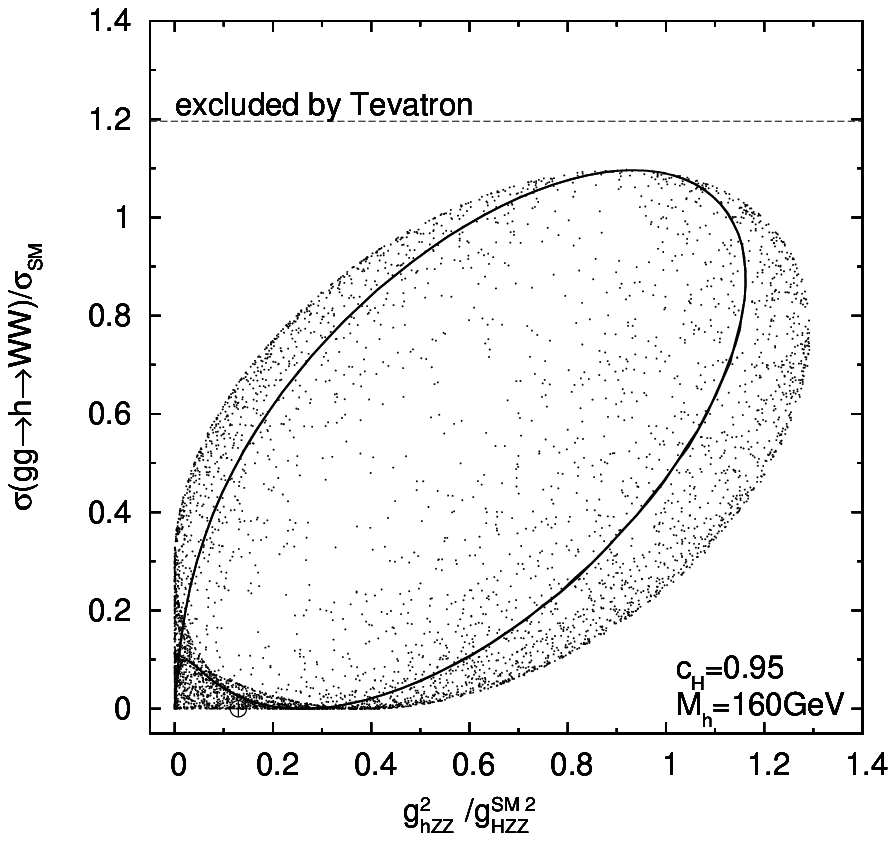}
\includegraphics{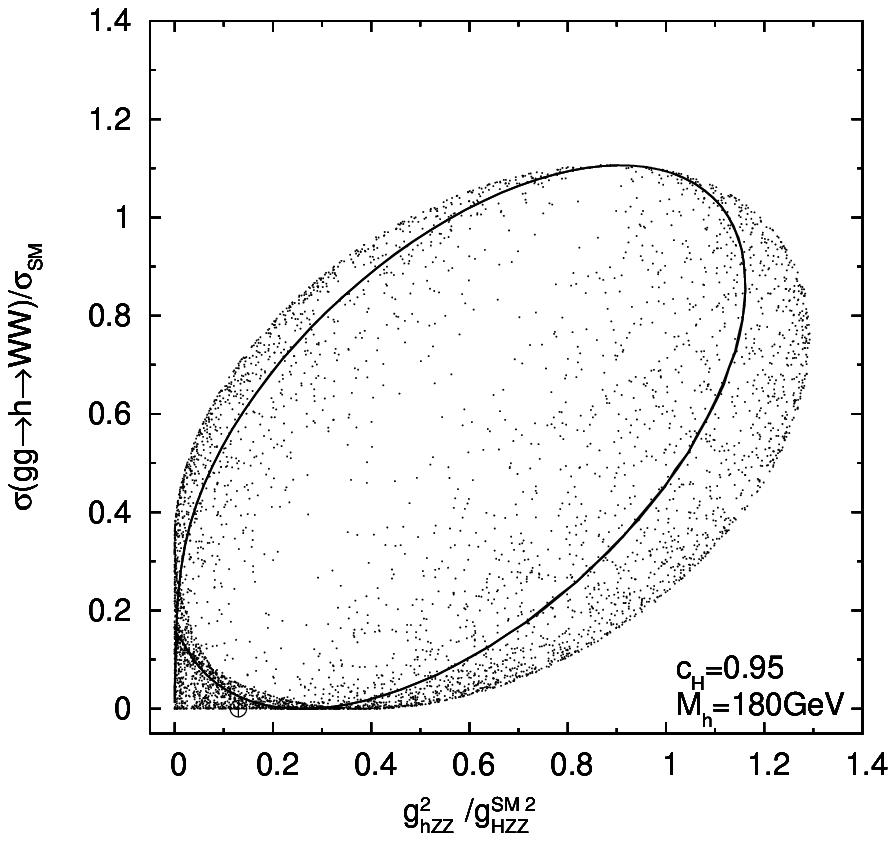}}
\caption{As in Fig.~\ref{fig:W9Z} but for $c_H = 0.95$.}
\label{fig:W95Z}
\end{figure}

\begin{figure}
\resizebox{\textwidth}{!}{
\includegraphics{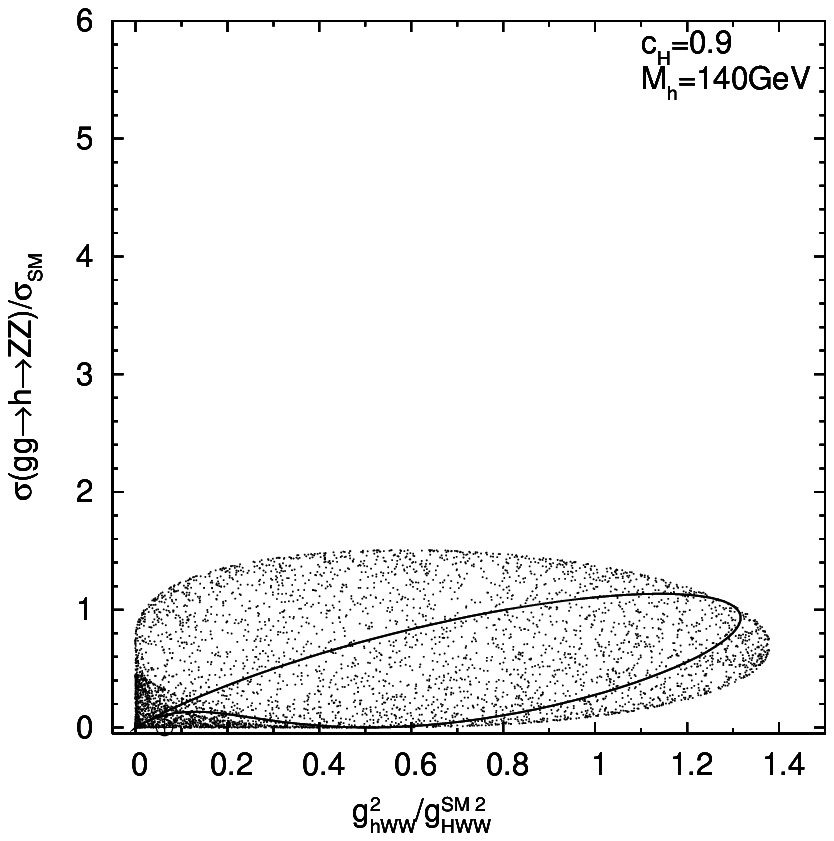}
\includegraphics{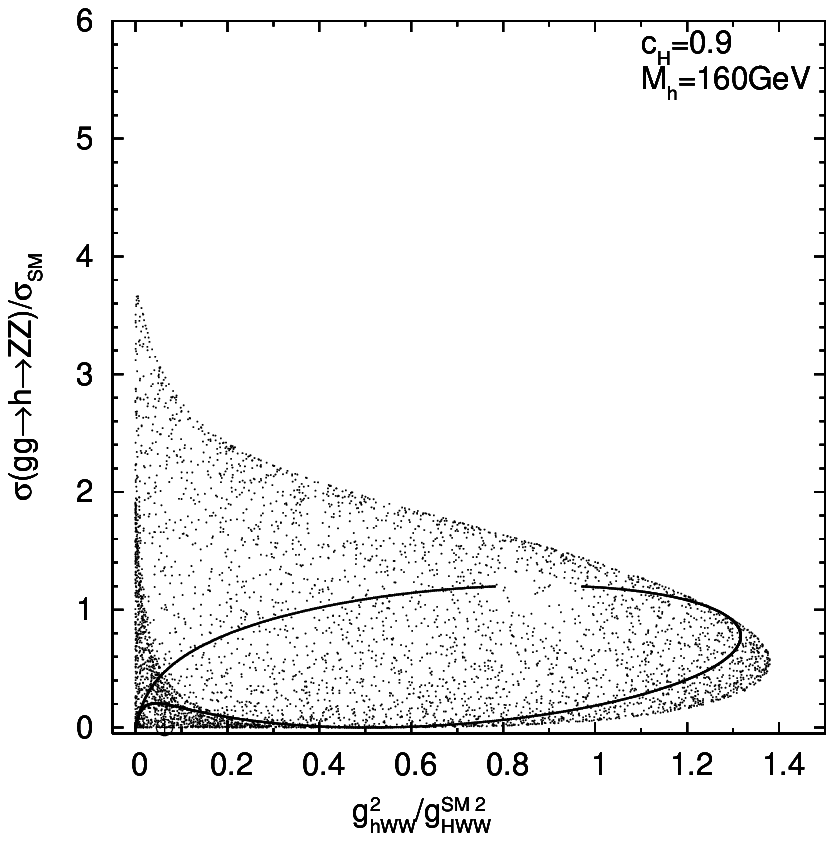}
\includegraphics{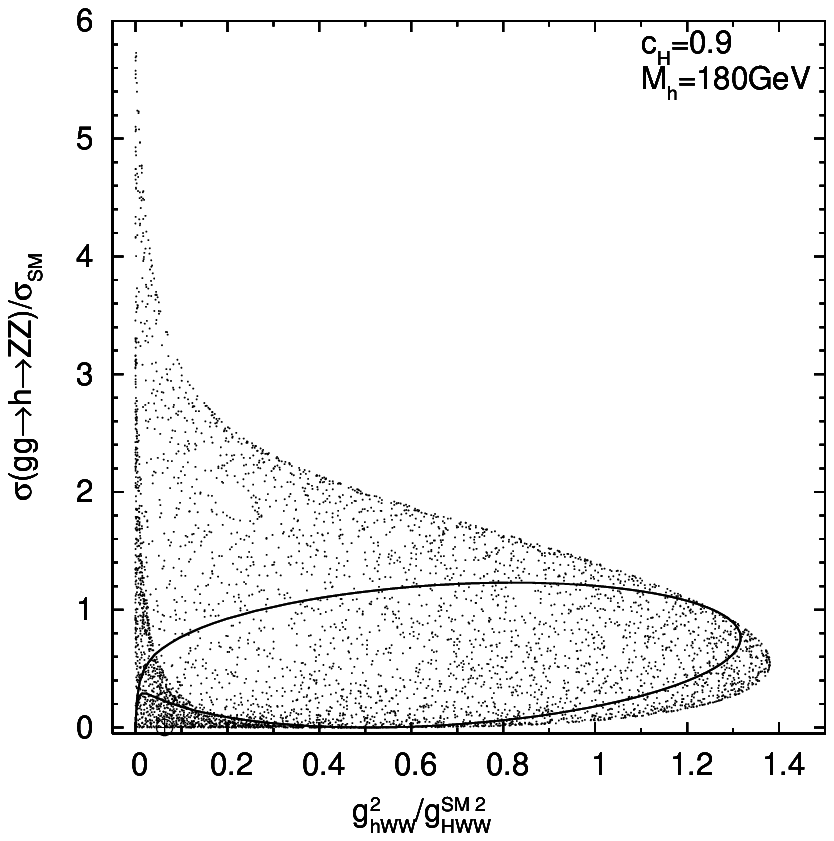}}
\caption{The rate for $gg \to h \to ZZ$ normalized to its SM value
  plotted as a function of $\bar g_{hWW}^2$, for $M_h = 140$, 160, and
  180~GeV (left to right), with $c_H = 0.9$.  The solid line shows
  points for which $h = \cos\theta H_1^0 + \sin\theta H_1^{0 \prime}$
  and the crossed circle near the origin indicates the point
  corresponding to $H_5^0$.  Points excluded by the Tevatron in
  Figs.~\ref{fig:W9W} and \ref{fig:W9Z} are not shown, leading to the
  gap in the solid line for $M_h = 160$~GeV.}
\label{fig:Z9W}
\end{figure}

\begin{figure}
\resizebox{\textwidth}{!}{
\includegraphics{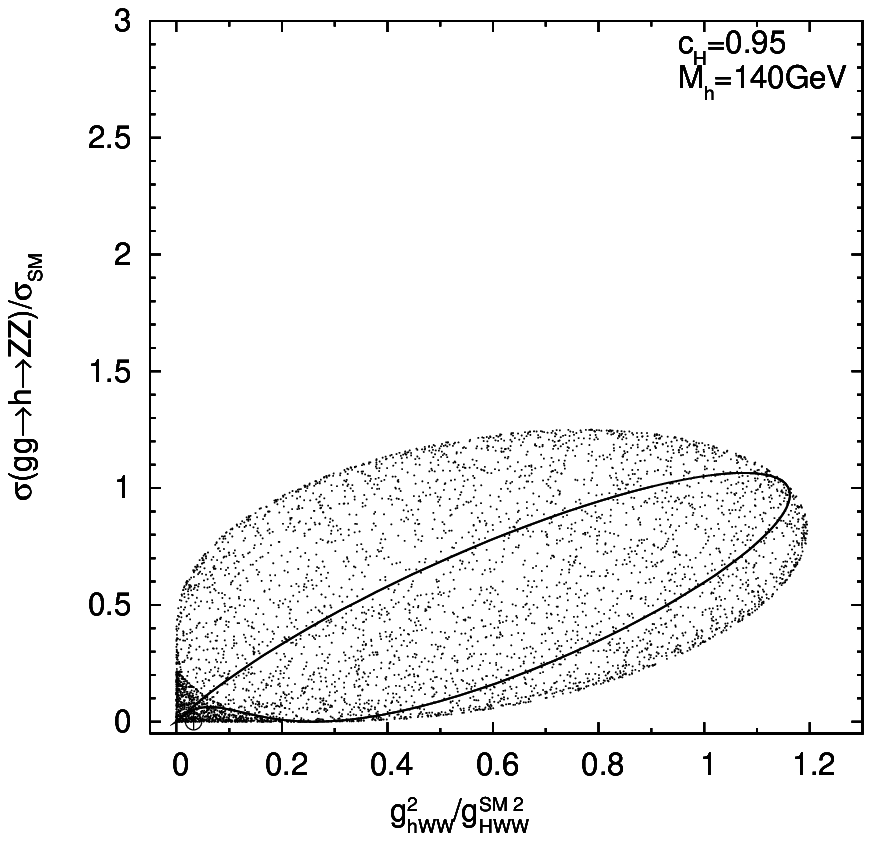}
\includegraphics{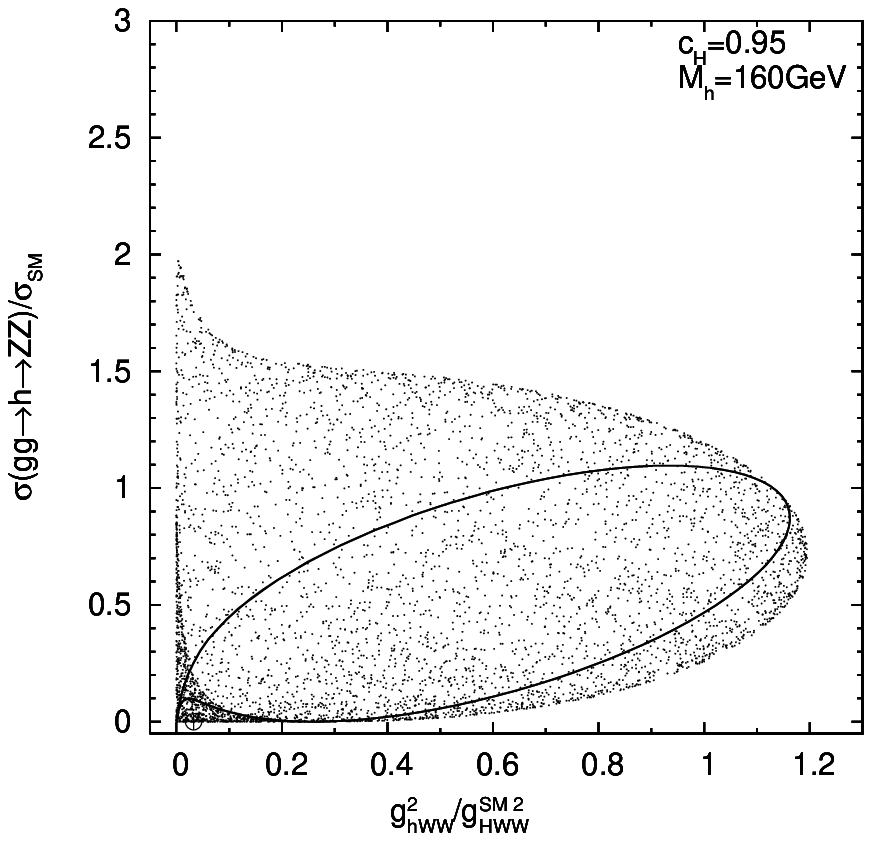}
\includegraphics{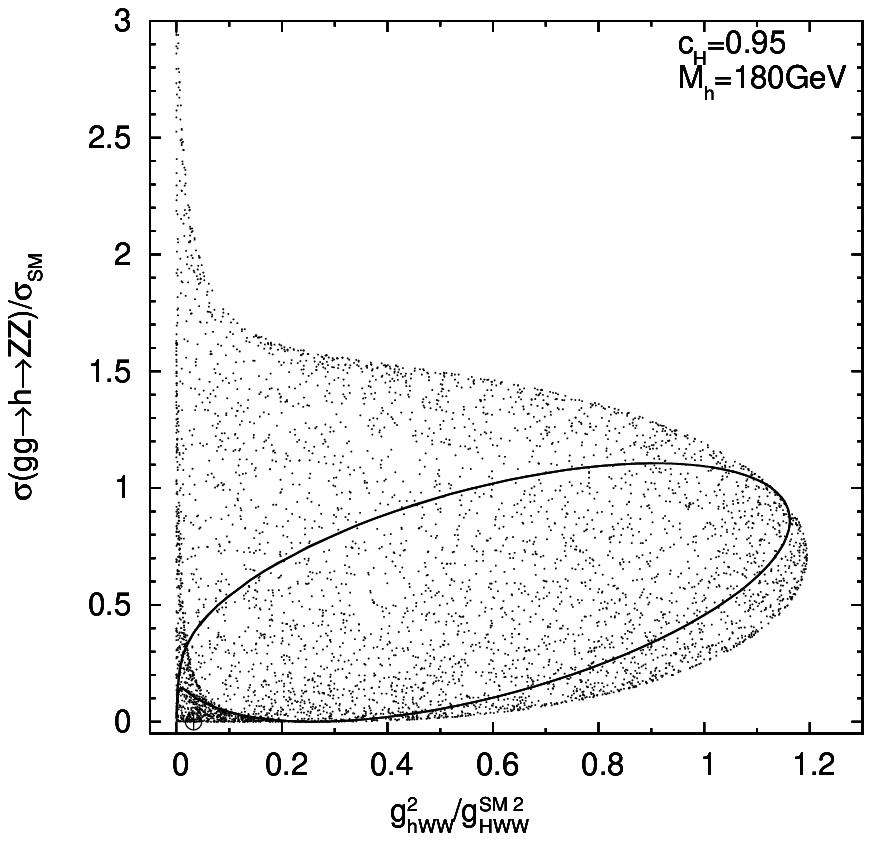}}
\caption{As in Fig.~\ref{fig:Z9W} but for $c_H = 0.95$.}
\label{fig:Z95W}
\end{figure}

\begin{figure}
\resizebox{\textwidth}{!}{
\includegraphics{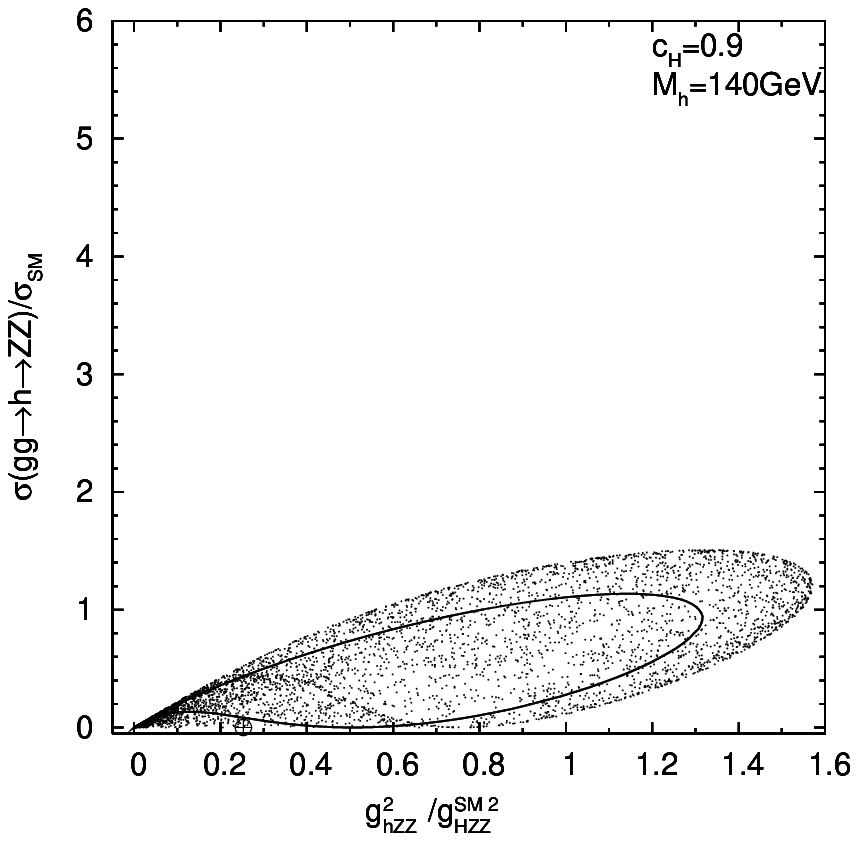}
\includegraphics{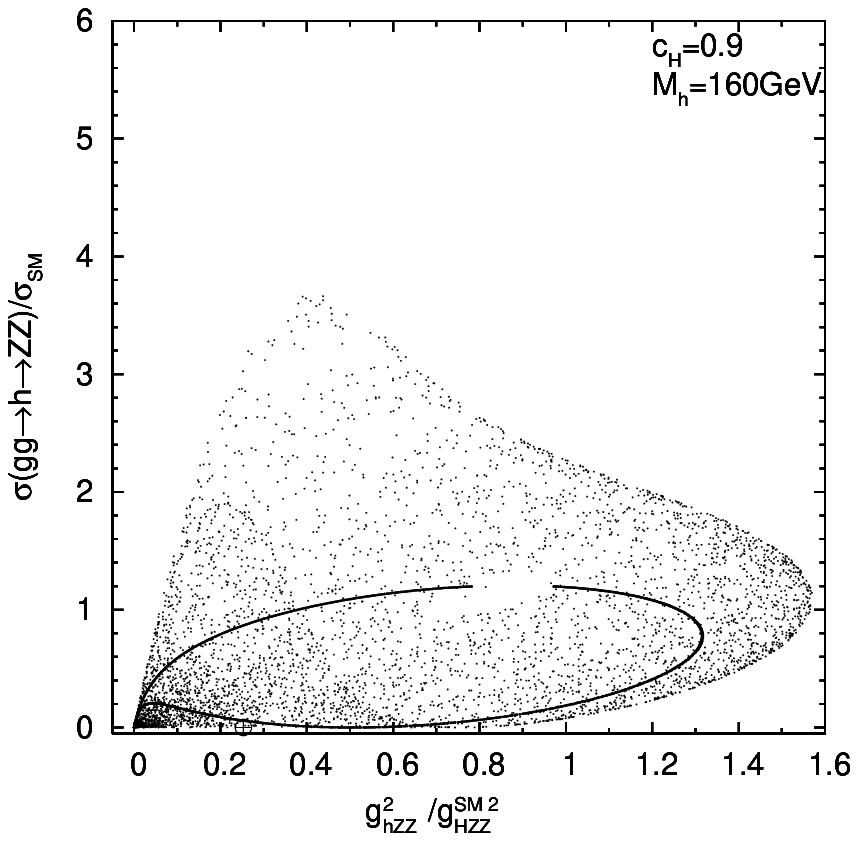}
\includegraphics{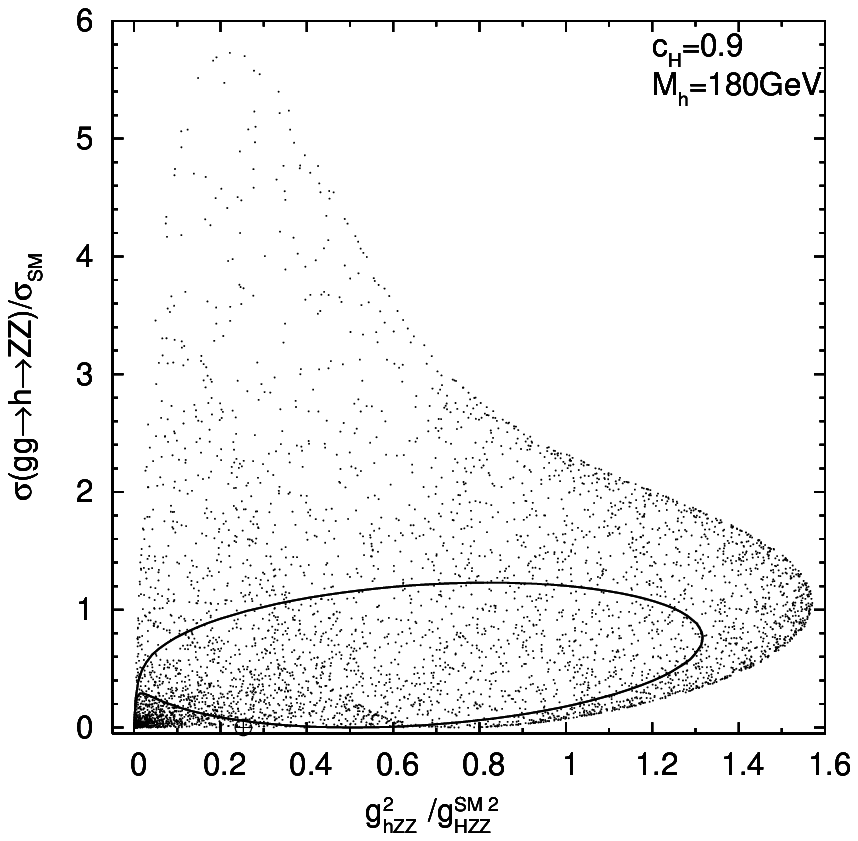}}
\caption{The rate for $gg \to h \to ZZ$ normalized to its SM value
  plotted as a function of $\bar g_{hZZ}^2$, for $M_h = 140$, 160, and
  180~GeV (left to right), with $c_H = 0.9$.  The solid line shows
  points for which $h = \cos\theta H_1^0 + \sin\theta H_1^{0 \prime}$
  and the crossed circle at the lower edge of the allowed region
  indicates the point corresponding to $H_5^0$.  Points excluded by
  the Tevatron in Figs.~\ref{fig:W9W} and \ref{fig:W9Z} are not shown,
  leading to the gap in the solid line for $M_h = 160$~GeV.}
\label{fig:Z9Z}
\end{figure}

\begin{figure}
\resizebox{\textwidth}{!}{
\includegraphics{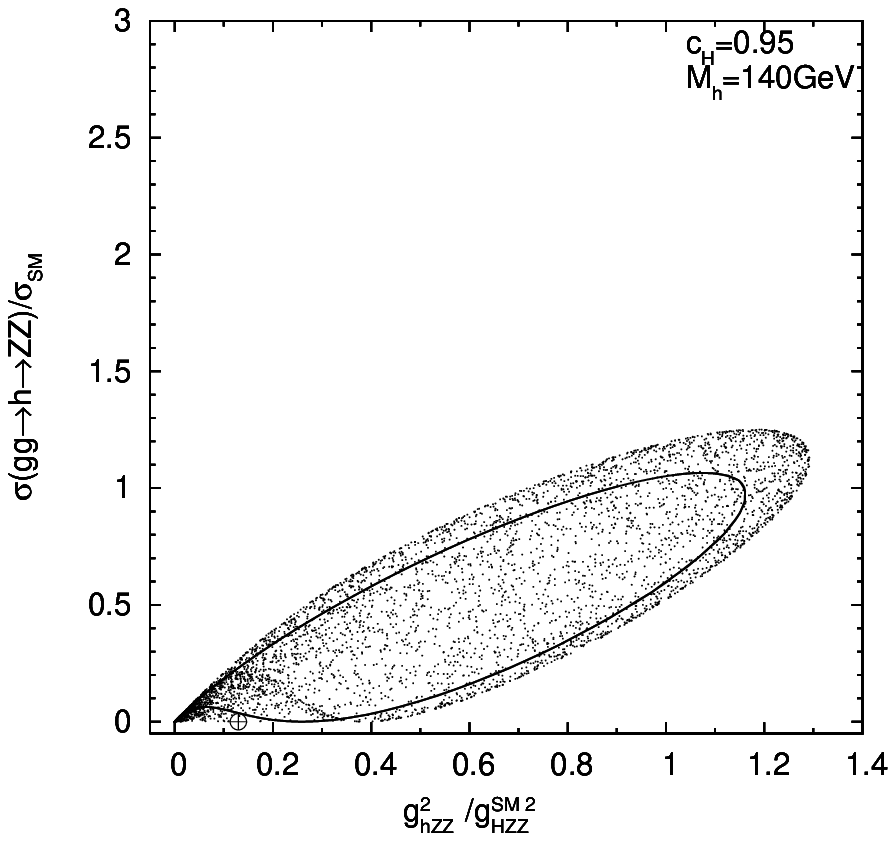}
\includegraphics{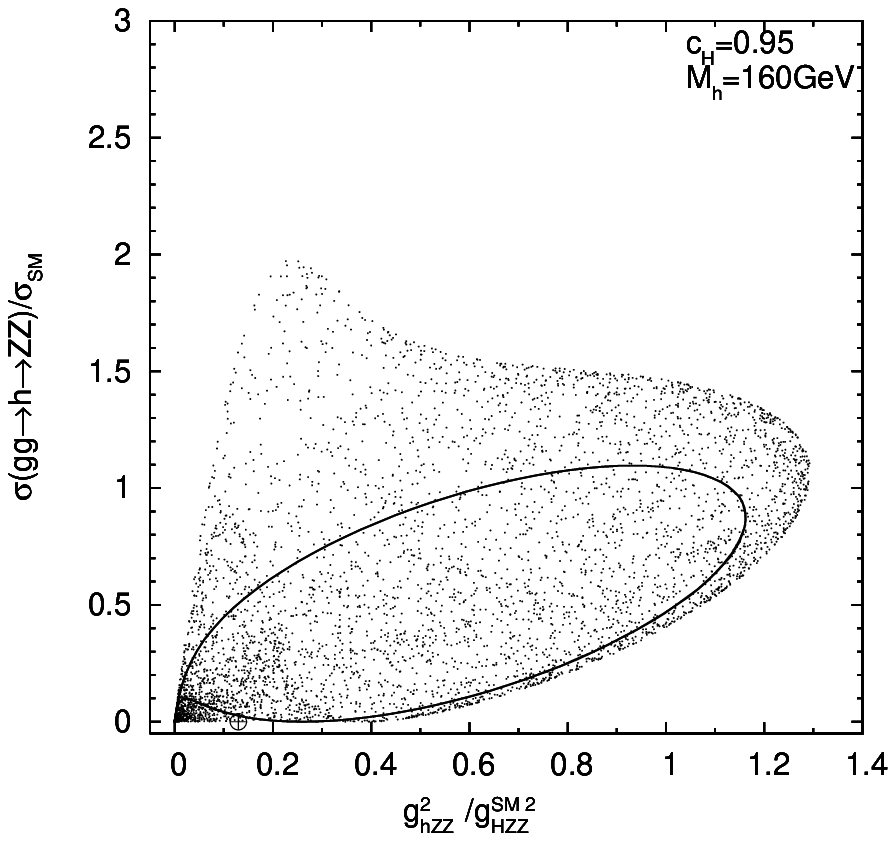}
\includegraphics{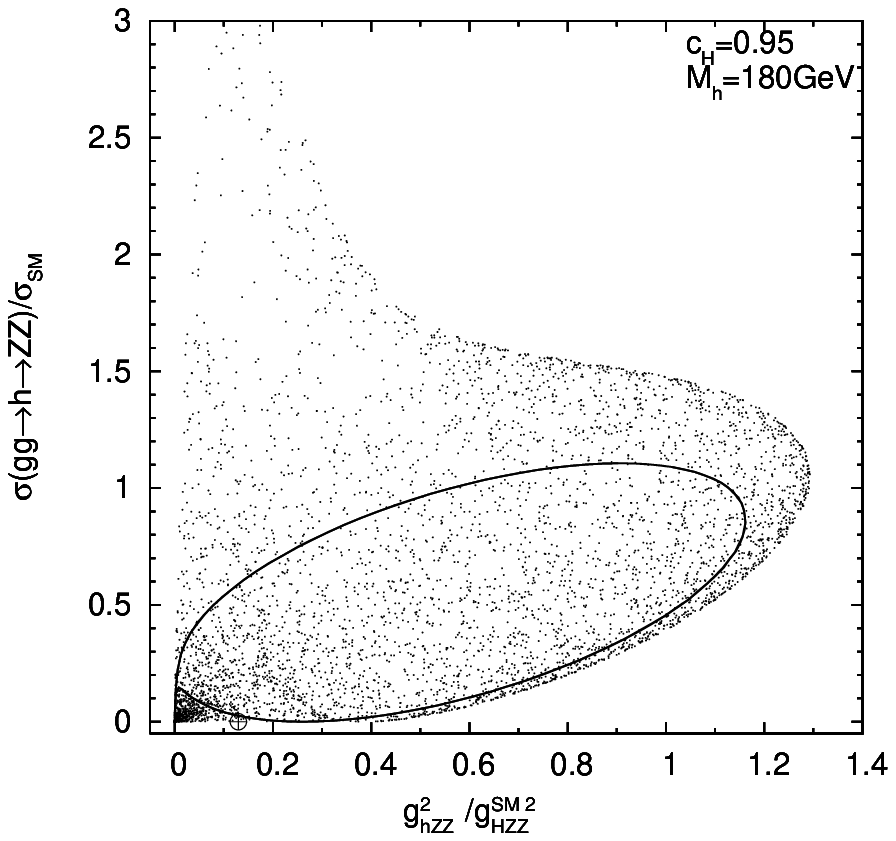}}
\caption{As in Fig.~\ref{fig:Z9Z} but for $c_H = 0.95$.}
\label{fig:Z95Z}
\end{figure}

We plot $\sigma(gg \to h \to WW)/\sigma_{\rm SM}$ in
Figs.~\ref{fig:W9W}--\ref{fig:W95Z} and $\sigma(gg \to h \to
ZZ)/\sigma_{\rm SM}$ in Figs.~\ref{fig:Z9W}--\ref{fig:Z95Z}.  We show
results for $M_h = 140$, 160, and 180~GeV and $c_H = 0.9$ and 0.95.
The points correspond to a general mixed state $h$, while the solid
line shows the custodial SU(2)-preserving mixture $h = \cos \theta
H_1^0 + \sin \theta H_1^{0 \prime}$.  For $h = H_5^0$, the $gg \to h$
cross section is zero because $H_5^0$ does not couple to fermions;
this is shown in the plots by a crossed circle.

%gg -> h -> WW plots

%gg -> h -> ZZ plots

%%%% end plots %%%%

The rate for $gg \to h \to WW$ can be enhanced by up to about 20\%
(10\%) compared to the corresponding SM rate for $c_H = 0.9$ (0.95),
as shown in Table~\ref{tab:ratesWW}.  The enhancement is due to an
interplay between the production and decay couplings.  For $c_H = 0.9$
(0.95), the $gg \to h$ production cross section is about $1.23 a^2$
($1.11 a^2$) times its SM value, for $a$ defined in Eq.~(\ref{eq:abc}).
The branching ratio for $h \to WW$ depends mainly on competition
between the $h \to WW$ and $h \to \bar b b$ partial widths, especially
for lower Higgs masses.

\begin{table}
\begin{tabular}{cccc}
\hline \hline
& & \multicolumn{2}{c}{Maximum of $\sigma(gg\to h\to WW)/\sigma_{\rm SM}$} \\
$c_H$ & ~$M_h$ [GeV]~ & ~~Overall~~ & $h \sim H_{1}^{0}, H_{1}^{0 \prime}$\\
\hline
 & 140 & 1.1581 & 1.1352 \\
0.9 & 160 & 1.2085 & 1.2056 \\
 & 180 & 1.2327 & 1.2296 \\
\hline
 & 140 & 1.0777 & 1.0656 \\
0.95 & 160 & 1.0971 & 1.0962 \\
 & 180 & 1.1069 & 1.1061 \\
\hline \hline
\end{tabular}
\caption{Maximum values of $\sigma(gg\to h\to WW)/\sigma_{\rm SM}$ for
  $c_H = 0.9$ and 0.95 and $M_h = 140$, 160, and 180~GeV.  The third
  column gives the maximum cross section enhancement obtainable when
  $h$ is any linear combination of $H_1^0$, $H_1^{0 \prime}$, and
  $H_5^0$, while the fourth column gives the maximum when $h$ is a
  linear combination of only $H_1^0$ and $H_1^{0 \prime}$.}
\label{tab:ratesWW}
\end{table}

For $c_H = 0.9$ and $M_h = 160$~GeV, the rate for $gg \to h \to WW$
can be enhanced enough to exceed the current limit on this cross
section from the Tevatron~\cite{Aaltonen:2010sv}.\footnote{Note that
  the Tevatron analysis in Ref.~\cite{Aaltonen:2010sv} selected
  specifically for the $gg \to H \to WW$ process, and is not a
  combination of multiple SM Higgs search channels.  The resulting
  cross section limits are thus directly applicable to the $gg \to h
  \to WW$ rates found here.}  Our calculation of the upper limit is
given in Table~\ref{tab:tevggww}.  Points excluded by the Tevatron are
not shown in the plots of $gg \to h \to ZZ$ in
Figs.~\ref{fig:Z9W}--\ref{fig:Z95Z}.

\begin{table}
\begin{tabular}{rccc}
\hline \hline 
& ~$M_h = 140$~GeV~ & ~160~GeV~ & ~180~GeV~ \\
\hline
$\sigma_{\rm SM}(gg \to H)$~\cite{Anastasiou:2008tj} &
   0.680~pb & 0.434~pb & 0.279~pb \\
BR$_{\rm SM}(H \to WW)$~\cite{HDECAY} &
   0.4916 & 0.9048 & 0.9325 \\
Tevatron upper limit on $\sigma(gg \to h \to WW)$~\cite{Aaltonen:2010sv}~ &
   1.29~pb & 0.47~pb & 0.41~pb \\
Upper limit on $\sigma(gg \to h \to WW)/\sigma_{\rm SM}$ &
   3.86 & 1.20 & 1.58 \\
\hline \hline
\end{tabular}
\caption{The $gg \to H$ cross section and $H \to WW$ branching
  fraction in the SM, Tevatron upper limit on the $gg \to h \to WW$
  cross section, and resulting upper limit on the cross section
  normalized to its SM value, for $M_h = 140$, 160, and 180~GeV.  The
  uncertainty on $\sigma_{\rm SM}(gg \to H)$ from the parton density
  functions is $\pm 8$--10\%~\cite{Anastasiou:2008tj}; here we have
  taken the cross section central value for our limit.}
\label{tab:tevggww}
\end{table}

The rate for $gg \to h \to ZZ$ can be enhanced much more dramatically,
by more than a factor of 5 (3) for $M_h = 180$ (160)~GeV and $c_H =
0.9$.  Details are given in Table~\ref{tab:ratesZZ}.  This large
enhancement occurs when the $hWW$ coupling is near zero but the $hZZ$
coupling is non-negligible, as can be seen by comparing
Figs.~\ref{fig:Z9W}--\ref{fig:Z95W} and
Figs.~\ref{fig:Z9Z}--\ref{fig:Z95Z}.  This suppresses the dominant
Higgs partial width to $WW$, resulting in BR($h \to ZZ$) close to 1
for the heavier masses.  Clearly, this can only happen if $\bar
g_{hWW} \neq \bar g_{hZZ}$, so it requires mixing between $H_5^0$ and
the custodial SU(2)-singlet states.  This would lead to an early LHC
discovery of such a Higgs in the ``golden mode,'' $h \to ZZ \to 4$
leptons, while the $h \to WW$ signal (which would provide the first
discovery in the SM) would be absent or very suppressed.

If $h = \cos\theta H_1^0 + \sin\theta H_1^{0 \prime}$, the enhancement
in the $ZZ$ channel is limited to about 20\% (10\%) for $c_H = 0.9$
(0.95).  Note that when $h = \cos\theta H_1^0 + \sin\theta H_1^{0
  \prime}$, the enhancement in the $WW$ and $ZZ$ channels is
identical, i.e., $\sigma(gg \to h \to WW)/\sigma_{\rm SM} = \sigma(gg
\to h \to ZZ)/\sigma_{\rm SM}$.  Such an enhancement would lead to a
marginally earlier discovery of $h$ in both channels at the LHC.  For
$c_H = 0.95$, however, the enhancement is about the same
size as the current uncertainty in $\sigma_{\rm SM}(gg \to H)$ from
the parton density functions~\cite{Anastasiou:2008tj}.

\begin{table}
\begin{tabular}{cccc}
\hline \hline
& & \multicolumn{2}{c}{Maximum of $\sigma(gg\to h\to ZZ)/\sigma_{\rm SM}$} \\
$c_H$ & ~$M_h$ [GeV]~ & ~~Overall~~ & $h \sim H_{1}^{0}, H_{1}^{0 \prime}$\\
\hline
 & 140 & 1.5068 & 1.1356 \\
0.9 & 160 & 3.6616 & 1.1978 \\
 & 180 & 5.7291 & 1.2297 \\
\hline
 & 140 & 1.2507 & 1.0659 \\
0.95 & 160 & 1.9716 & 1.0962 \\
 & 180 & 3.0156 & 1.1061 \\
\hline \hline
\end{tabular}
\caption{As in Table~\ref{tab:ratesWW} but for 
$\sigma(gg \to h \to ZZ)/\sigma_{\rm SM}$.}
\label{tab:ratesZZ}
\end{table}

%=======================================================================
\section{Higgs coupling extraction and the triplet nature of $h$}
\label{sec:howtotell}

In models containing only Higgs doublets and singlets, the $hVV$ couplings
satisfy
\begin{equation}
  \bar g_{hWW} = \bar g_{hZZ} \equiv \bar g_{hVV}, 
  \qquad |\bar g_{hVV}| \leq 1.
\end{equation}
These relations are assumed to hold in the Higgs coupling fits of
Refs.~\cite{Duhrssen-ATLAS,Duhrssen:2004cv,Lafaye:2009vr}; however,
both of them are violated in the Georgi-Machacek model.

The relation $\bar g_{hWW} = \bar g_{hZZ}$ is easy to test by taking
the ratio of any two Higgs signal rates with the same production
mechanism and decays to $WW$ and $ZZ$, respectively:
\begin{equation}
  \frac{\sigma(X \to h \to WW)/\sigma_{\rm SM}(X \to h \to WW)}
       {\sigma(X \to h \to ZZ)/\sigma_{\rm SM}(X \to h \to ZZ)} 
  = \frac{\bar g_{hWW}^2}{\bar g_{hZZ}^2}.
\end{equation}

Determining whether $\bar g_{hVV}$ is greater than 1 is more
complicated, and is not always possible at the LHC.  The approach is
as follows.  One of the key ingredients in the Higgs coupling fits is
the measurement of the rate for Higgs production in vector boson
fusion ($WW$, $ZZ \to h$) followed by decay to $WW$.  This rate can be
written as
\begin{equation}
  \sigma({\rm VBF} \to h \to WW) = \sigma({\rm VBF} \to h) \,
  {\rm BR}(h \to WW).
  \label{eq:extraction}
\end{equation}
The usual Higgs coupling fit involves assuming that $\sigma({\rm VBF}
\to h)$ is no larger than its SM value, i.e., $\sigma^{\rm max}({\rm
  VBF} \to h) = \sigma_{\rm SM}({\rm VBF} \to H)$.  Together with the
rate measurement, this sets a lower bound on BR($h \to WW$):
\begin{equation}
  {\rm BR}^{\rm min}(h \to WW) = \frac{\sigma({\rm VBF} \to h \to WW)}
  {\sigma^{\rm max}({\rm VBF} \to h)}.
  \label{eq:BRmin}
\end{equation}
If the assumption that $\sigma({\rm VBF} \to h) \leq \sigma_{\rm
  SM}({\rm VBF} \to H)$ is false, the resulting value of BR$^{\rm
  min}(h \to WW)$ can be greater than one, which is clearly
unphysical.  In particular, for $c_H = 0.9$ (0.95), we find that
$\sigma({\rm VBF} \to h)$ can be as much as 31\% (16\%)
larger than in the SM.  An unphysical BR$^{\rm min}(h \to WW)$
would then be obtained using Eq.~(\ref{eq:BRmin}) if the true value of
BR($h \to WW$) is greater than 0.76 (0.86), which is true in the SM
for $M_H$ in the range 154--192~GeV (159--184~GeV).  

This range can be extended by considering Higgs decays to other final
states as follows.  If the Higgs is detectable in the $WW$ final state
and some other final state ($ZZ$, $\tau\tau$, etc.) via a common
production mechanism, we can write
\begin{equation}
  \frac{{\rm BR}(h \to ZZ)}{{\rm BR}(h \to WW)} 
  = \frac{\sigma(X \to h \to ZZ)}{\sigma(X \to h \to WW)}, 
  \qquad
  \frac{{\rm BR}(h \to \tau\tau)}{{\rm BR}(h \to WW)}
  = \frac{\sigma(X \to h \to \tau\tau)}{\sigma(X \to h \to WW)},
\end{equation}
etc.  The total branching fraction of the Higgs to detected modes is
then given by
\begin{equation}
  {\rm BR}(h \to {\rm detected}) = {\rm BR}(h \to WW)
  \left[1 + \frac{{\rm BR}(h \to ZZ)}{{\rm BR}(h \to WW)}
    + \frac{{\rm BR}(h \to \tau\tau)}{{\rm BR}(h \to WW)}
    + \cdots \right].
\end{equation}
If inserting BR$^{\rm min}(h \to WW)$ from Eq.~(\ref{eq:BRmin}) on the
right-hand side yields a value of BR($h \to {\rm detected}$) greater
than 1, then we can conclude that the assumption that $\sigma({\rm
  VBF} \to h) \leq \sigma_{\rm SM}({\rm VBF} \to H)$ is false.  

If the detectable branching fraction of the Higgs is sufficiently
small, the technique discussed above will fail.  In that case,
detection of the triplet nature of $h$ at the LHC would rely on the
discovery of additional Higgs states.  Search prospects at the LHC for
single production of $\chi^{++}$ with decays to $W^+W^+$ were studied
in Refs.~\cite{Azuelos:2004dm,Godfrey:2010qb}, and for pair production
of $\chi^{++}\chi^{--}$ again with decays to like-sign $W$ bosons in
Ref.~\cite{Han:2007bk}.\footnote{Reference~\cite{Han:2007bk} also
  considered $\chi^{++}$ decays to like-sign charged leptons via
  lepton-number-violating Yukawa couplings that could be responsible
  for neutrino masses.  For the triplet vevs that we consider here,
  $v_{\chi} = 54$ (38)~GeV for $c_H = 0.9$ (0.95), such Yukawa
  couplings would be of order $10^{-11}$ and dilepton decays would be
  totally negligible.}  Single production of $H_5^+ \equiv (\chi^+ -
\xi^+)/\sqrt{2}$ through $W^+Z$ fusion with decays back to $W^+Z$ was
also studied in Refs.~\cite{Asakawa:2006gm,Godfrey:2010qb}.

We note that the technique discussed here can yield only a lower
bound on $\sigma({\rm VBF} \to h)$.  Furthermore, removing the SM
assumption that $\sigma({\rm VBF} \to h) \leq \sigma_{\rm SM}({\rm
  VBF} \to H)$ reopens the parameter degeneracy in the Higgs coupling
fit discussed in the introduction.  While ratios of Higgs couplings
would still be measurable at the LHC, knowledge of absolute Higgs
couplings would have to wait for direct model-independent measurements
at an $e^+e^-$ collider~\cite{ILCHiggs}.

%=======================================================================
\section{Conclusions}
\label{sec:conclusions}

We studied the couplings of a CP-even neutral Higgs state $h$ in the
Georgi-Machacek model with Higgs triplets.  We found that the effect
of the triplet components of $h$ on its couplings to $WW$ and $ZZ$ can
be significant, even for relatively small triplet vevs consistent with
the experimental upper bounds on the top quark Yukawa coupling.

If $h$ is a mixture of only custodial SU(2) singlets, the $hWW$ and
$hZZ$ couplings can be enhanced by up to 15\% (8\%) for $c_H = 0.9$
(0.95), but the ratio of these couplings is the same as in the SM.
If $h$ is a generic mixture of the CP-even neutral states, the
ratio of its $WW$ and $ZZ$ couplings can be very different than in the
SM; in fact, one of these couplings can be zero while the other is
finite.  Such a suppression of the $hWW$ coupling while the $hZZ$
coupling is nonzero can lead to large enhancements in the rate for $gg
\to h \to ZZ$, especially for $M_h \gtrsim 160$~GeV where the SM Higgs
width is dominated by decays to $WW$.

While a ratio of $hWW$ and $hZZ$ couplings different than in the SM
would be easy to measure by taking ratios of rates in these channels,
an overall enhancement in the $hVV$ coupling could only be detected at
the LHC if it leads to a rate in vector boson fusion with decays to
$WW$ too large to be consistent with the SM assumption for this
coupling.

%%%%%%%%%%%%%%%%%%%%%%%%%%%%%%%%%%%%%%%%%%%%%%%%%%%%%%%%%%%%%%%%%%%%%%%%%%

\begin{acknowledgments}
We thank K.~Moats for helpful comments on the manuscript.
This work was supported by the Natural Sciences and Engineering
Research Council of Canada.  
\end{acknowledgments}

%%%%%%%%%%%%%%%%%%%%%%%%%%%%%%%%%%%%%%%%%%%%%%%%%%%%%%%%%%%%%%%%%%%%%%%%%%

\appendix

\section{Covariant derivative and SU(2) generators for triplets}
\label{app:covar}

The gauge-covariant derivative in Eq.~(\ref{eq:kineticterms}) is given by 
\begin{equation}
  \mathcal{D}_{\mu} = \partial_{\mu} + i g T^a W^a_{\mu} 
  + i g^{\prime} Y B_{\mu}.
\end{equation}
The SU(2) generators for a doublet representation are given by
$T^a = \sigma^a/2$, where $\sigma^a$ are the 2$\times$2 Pauli matrices.
The SU(2) generators for a triplet representation are
\begin{equation}
  T^1 = \frac{1}{\sqrt{2}} \left( \begin{array}{ccc} 
    0~ & 1~ & 0 \\
    1~ & 0~ & 1 \\
    0~ & 1~ & 0 \end{array} \right), \qquad
  T^2 = \frac{1}{\sqrt{2}} \left( \begin{array}{ccc}
    0~ & -i & 0 \\
    i~ & 0  & -i \\
    0~ & i & 0 \end{array} \right), \qquad
  T^3 = \left( \begin{array}{ccc}
    1~ & 0 & 0 \\
    0~ & 0 & 0 \\
    0~ & 0 & -1 \end{array} \right).
\end{equation}

%==================================================================
\section{Scalar potential and mass matrix}
\label{app:potential}

Reference~\cite{Gunion:1990dt} presented the most general
gauge-invariant scalar potential for this model containing only terms
with an even number of fields.  While not completely general, this
restricted potential captures most of the
physics~\cite{Chanowitz:1985ug}.  In our notation it is,
\begin{eqnarray}
  V_{\rm even} &=& \mu_1^2 \Phi^{\dagger} \Phi + \mu_2^2 X^{\dagger} X
  + \mu_3^2 \Xi^{\dagger} \Xi \nonumber \\ 
  && + \lambda_1 (\Phi^{\dagger} \Phi)^2 + \lambda_2 | X^T C X |^2 
  + \lambda_3 ( \Phi^{\dagger} T^a \Phi ) ( X^{\dagger} T^a X ) \nonumber \\ 
  && + \lambda_4 \left[ (\Phi^{\dagger} T^a \Phi^c) (\Xi^{\dagger} T^a X) 
    + {\rm h.c.} \right] 
  + \lambda_5 (\Phi^{\dagger} \Phi) (X^{\dagger} X) 
  + \lambda_6 (\Phi^{\dagger} \Phi) (\Xi^{\dagger} \Xi) \nonumber \\ 
  && + \lambda_7 (X^{\dagger} X)^2 + \lambda_8 (\Xi^{\dagger} \Xi)^2 
  + \lambda_9 |X^{\dagger} \Xi|^2 
  + \lambda_{10} (\Xi^{\dagger} \Xi) (X^{\dagger} X),
\end{eqnarray}
where $\lambda_i$ are dimensionless and $\mu_i^2$ have dimension mass
squared.  Here $\Phi^c = i \sigma_2 \Phi^*$ and
\begin{equation}
  C = \left( \begin{array}{ccc}
    0~ & 0 & ~1 \\
    0~ & -1 & ~0 \\
    1~ & 0 & ~0 \end{array} \right)
\end{equation}
is the conjugation operator for the triplet fields such that
$X^c = C X^*$ transforms as a triplet of SU(2).

We add the most general gauge-invariant set of trilinear terms,
\begin{equation}
  V_{\rm trilinear} = \kappa_1 [(\Phi^{\dagger} T^a \Phi^c) X^a + {\rm h.c.}]
  + \kappa_2 (\Phi^{\dagger} T^a \Phi) \Xi^a
  + \kappa_3 X^{* a} (i \epsilon^{abc} X^b \Xi^c),
\end{equation}
where $\kappa_i$ have dimensions of mass and $\epsilon^{abc}$ is the 
totally antisymmetric tensor.  

Minimizing the potential allows us to eliminate $\mu_i^2$ in favor
of the scalar vevs:
\begin{eqnarray}
  \mu_1^2 &=& - \kappa_1 v_{\chi} + \kappa_2 \frac{v_{\xi}}{2}
  - \lambda_1 v_{\phi}^2 
  - \left( \frac{\lambda_3}{4} + \frac{\lambda_5}{2} \right) v_{\chi}^2
  - \lambda_4 v_{\chi} v_{\xi} - \lambda_6 v_{\xi}^2 \nonumber \\
  \mu_2^2 &=& - \kappa_1 \frac{v_{\phi}^2}{2 v_{\chi}} + \kappa_3 v_{\xi}
  - \left( \frac{\lambda_3}{4} + \frac{\lambda_5}{2} \right) v_{\phi}^2
  - \lambda_4 \frac{v_{\phi}^2 v_{\xi}}{2 v_{\chi}}
  - \lambda_7 v_{\chi}^2 - \lambda_{10} v_{\xi}^2 \nonumber \\
  \mu_3^2 &=& \kappa_2 \frac{v_{\phi}^2}{8 v_{\xi}}
  + \kappa_3 \frac{v_{\chi}^2}{4 v_{\xi}} 
  - \lambda_4 \frac{v_{\phi}^2 v_{\chi}}{4 v_{\xi}}
  - \lambda_6 \frac{v_{\phi}^2}{2} 
  - 2 \lambda_8 v_{\xi}^2 - \lambda_{10} \frac{v_{\chi}^2}{2}.
\end{eqnarray}
Applying these minimization conditions, the quadratic terms in the potential
involving CP-even neutral fields are,
\begin{eqnarray}
  V_{m^2 \ {\rm terms}} &=& 
  (\phi^{0,r})^2 \left[ \lambda_1 v_{\phi}^2 \right] \nonumber \\
  && + (\chi^{0,r})^2 \left[ -\kappa_1 \frac{v_{\phi}^2}{4 v_{\chi}}
    - \lambda_4 \frac{v_{\phi}^2 v_{\xi}}{4 v_{\chi}}
    + \lambda_7 v_{\chi}^2 \right] \nonumber \\
  && + (\xi^0)^2 \left[ \kappa_2 \frac{v_{\phi}^2}{8 v_{\xi}}
    + \kappa_3 \frac{v_{\chi}^2}{4 v_{\xi}}
    - \lambda_4 \frac{v_{\phi}^2 v_{\chi}}{4 v_{\xi}}
    + 4 \lambda_8 v_{\xi}^2 \right] \nonumber \\
  && + \phi^{0,r} \chi^{0,r} \left[ \kappa_1 v_{\phi}
    + \left( \frac{\lambda_3}{2} + \lambda_5 \right) v_{\phi} v_{\chi}
    + \lambda_4 v_{\phi} v_{\xi} \right] \nonumber \\
  && + \phi^{0,r} \xi^0 \left[ -\kappa_2 \frac{v_{\phi}}{2}
    + \lambda_4 v_{\phi} v_{\chi} + 2 \lambda_6 v_{\phi} v_{\xi} \right]
  \nonumber \\
  && + \chi^{0,r} \xi^0 \left[ -\kappa_3 v_{\chi}
    + \frac{\lambda_4}{2} v_{\phi}^2 + 2 \lambda_{10} v_{\chi} v_{\xi}
    \right].
\end{eqnarray}
Rewriting the fields in the $H_1^0$, $H_1^{0 \prime}$, $H_5^0$ basis
given in Eq.~(\ref{eq:H1H5}), using the notation of
Eq.~(\ref{eq:vevs}) for the vevs, and setting $\Delta \rho = 0$, we
obtain
\begin{eqnarray}
  V_{m^2 \ {\rm terms}} &=& v_{\rm SM}^2 \left\{
  (H_1^0)^2 \lambda_1 c_H^2 \rule[-5pt]{0mm}{22pt} \right. \nonumber \\
  && + (H_1^{0 \prime})^2 
  \left[ -\frac{(2 \sqrt{2} \kappa_1 - \kappa_2)}{v_{\rm SM}}
    \frac{c_H^2}{6 \sqrt{2} s_H}
    - \frac{\kappa_3}{v_{\rm SM}} \frac{s_H}{4 \sqrt{2}} 
    + \lambda_4 \frac{c_H^2}{3 \sqrt{2}}
    + (\lambda_7 + \lambda_8 + \lambda_{10}) \frac{s_H^2}{6}
    \right] 
  \nonumber \\
  && + (H_5^0)^2 \left[ -\frac{(\kappa_1 - \sqrt{2} \kappa_2)}{v_{\rm SM}}
    \frac{c_H^2}{6 s_H}
     + \frac{\kappa_3}{v_{\rm SM}} \frac{s_H}{2 \sqrt{2}}
    - \lambda_4 \frac{3 c_H^2}{4 \sqrt{2}}
    + (\lambda_7 + 4 \lambda_8 - 2 \lambda_{10}) \frac{s_H^2}{12}
    \right] \nonumber \\
  && + H_1^0 H_1^{0 \prime} \left[ 
    \frac{(2 \sqrt{2} \kappa_1 - \kappa_2)}{v_{\rm SM}} \frac{c_H}{2 \sqrt{3}}
    + (\lambda_3 + 2 \sqrt{2} \lambda_4 + 2 \lambda_5 + 2 \lambda_6) 
    \frac{c_H s_H}{2 \sqrt{6}}
    \right] \nonumber \\
  && + H_1^0 H_5^0 \left[ 
    - \frac{(\sqrt{2} \kappa_1 + \kappa_2)}{v_{\rm SM}} \frac{c_H}{\sqrt{6}} 
   + ( -\lambda_3 + \sqrt{2} \lambda_4 - 2 \lambda_5 + 4 \lambda_6) 
   \frac{c_H s_H}{4 \sqrt{3}}
    \right] \nonumber \\
  && \left. + H_1^{0 \prime} H_5^0 \left[ 
    \frac{(\sqrt{2} \kappa_1 + \kappa_2)}{v_{\rm SM}} \frac{c_H^2}{3 s_H}
    + (-2 \lambda_7 + 4 \lambda_8 + \lambda_{10}) \frac{s_H^2}{6 \sqrt{2}}
    \right] \right\}.
\end{eqnarray}
The last two terms, representing mixing between $H_1^0$ and $H_5^0$
and $H_1^{0 \prime}$ and $H_5^0$, respectively, violate the custodial
SU(2).  Imposing the conditions given in Ref.~\cite{Gunion:1990dt} for
the scalar potential to be invariant under a global SU(2)$_R$ causes
the $\lambda_i$ contributions of these last two terms to
vanish.\footnote{The relevant conditions are $\lambda_3 = \sqrt{2}
  \lambda_4$, $\lambda_5 = 2 \lambda_6$, $2 \lambda_7 = \lambda_9 + 2
  \lambda_{10}$, and $4 \lambda_8 = \lambda_9 + \lambda_{10}$.}  When
$\kappa_2 = -\sqrt{2} \kappa_1$ the remaining contribution vanishes as
well.

Even if the custodial SU(2)--violating couplings are small, the mixing
between the singlets and $H_5^0$ can be large if their masses are 
degenerate enough.\footnote{The near-degenerate case was studied in
a model with one doublet and one triplet in Ref.~\cite{Akeroyd}.}

%==========================

\end{document}